\documentclass[journal,romanappendices,onecolumn,draftclsnofoot]{IEEEtran}

\usepackage{graphicx,epsf}
\usepackage{amsfonts,amsthm,amsmath,bbm,xspace}
\usepackage{tabularx,ragged2e}

\def\Rset{\mathbb{R}}
\def\Nset{\mathbb{N}}

\def\opL{\mathcal{L}}
\def\opK{\mathcal{K}}
\def\opI{\mathcal{I}}

\def\Dkl{D_{\mathrm{kl}}}
\def\Hf{H_\phi}
\def\Df{D_\phi}
\def\Jf{J^{(\phi)}}

\def\lr{\frac{p_1}{p_0}}
\def\lrd{p_1\|p_0}

\def\Tr{\mbox{Tr}}
\def\Hess{\mathcal{H}}
\def\un{\mathbbm{1}}
\def\div{\operatorname{div}}
\def\;{\, ; \,}

\def\Esp{\mathbb{E}}
\def\MSE{\operatorname{MSE}}
\def\MMSE{\operatorname{MMSE}}

\def\ie{{\it i.e.,}\xspace}
\def\eg{{\it e.g.,}\xspace}

\newtheorem{definition}{Definition}
\newtheorem{proposition}{Proposition}

\newtheorem*{theorem*}{Theorem}
\newtheorem*{lemma*}{Lemma}

\title{Generalization of  the de  Bruijn's identity to  general $\phi$-entropies
  and $\phi$-Fisher informations}

\author{Irene~Valero~Toranzo\thanks{I.  V.  Toranzo is with  both the GIPSA-Lab,
    Image and  Signal Processing Departement, 11 rue  des Math\'ematiques, 38402
    St  Martin d'H\`eres,  France and  the Departamento  de  F\'isica At\'omica,
    Molecular   y  Nuclear,   Universidad  de   Granada,   18071-Granada,  Spain
    (ivtoranzo@ugr.es)}, Steeve~Zozor and Jean-Marc~Brossier\thanks{S. Zozor and
    J.-M.   Brossier  are  with  the  GIPSA-Lab,  Image  and  Signal  Processing
    Departement, 11  rue des Math\'ematiques, 38402 St  Martin d'H\`eres, France
    (steeve.zozor@gipsa-lab.grenoble-inp.fr;
    jean-marc.brossier@gipsa-lab.grenoble-inp.fr)}}

\begin{document}
\maketitle

\begin{abstract}
  In this paper, we propose  generalizations of the de Bruijn's identities based
  on extensions of the Shannon  entropy, Fisher information and their associated
  divergences or relative measures.  The foundation of these generalizations are
  the $\phi$-entropies and divergences  of the Csisz\'ar's class (or Salicr\'u's
  class)   considered   within   a   multidimensional  context,   included   the
  monodimensional case, and for several  type of noisy channels characterized by
  a  more  general  probability  distribution  beyond  the  well-known  Gaussian
  noise. It is found that the  gradient and/or the hessian of these entropies or
  divergences  with respect  to  the  noise parameters  give  naturally rise  to
  generalized versions of the Fisher  information or divergence, which are named
  as the $\phi$-Fisher information  (divergence). The obtained identities can be
  viewed    as   further    extensions    of   the    classical   de    Bruijn's
  identity. Analogously, it  is shown that a similar  relation holds between the
  $\phi$-divergence and  a extended mean-square error,  named $\phi$-mean square
  error, for the Gaussian channel.
\end{abstract}


\section{Introduction}
\label{Intro:sec}

The  goal of this  paper is  to extend  the de  Bruijn's identity,  relating two
quantities of information, namely the differential Shannon entropy of the output
of  a  Gaussian channel,  and  its  Fisher  information~\cite{Sta59}. These  two
quantities  are  very  important   in  information  theory,  in  statistics,  in
statistical  physics  and  in  signal processing~\cite{CovTho06,  Fri04,  Kay93,
  VigBer03, Bas89, Bas13, HerMa02, DarWue00, EbeMol00, DehLop10:12}.

\begin{figure}
\centerline{\includegraphics[width=6cm]{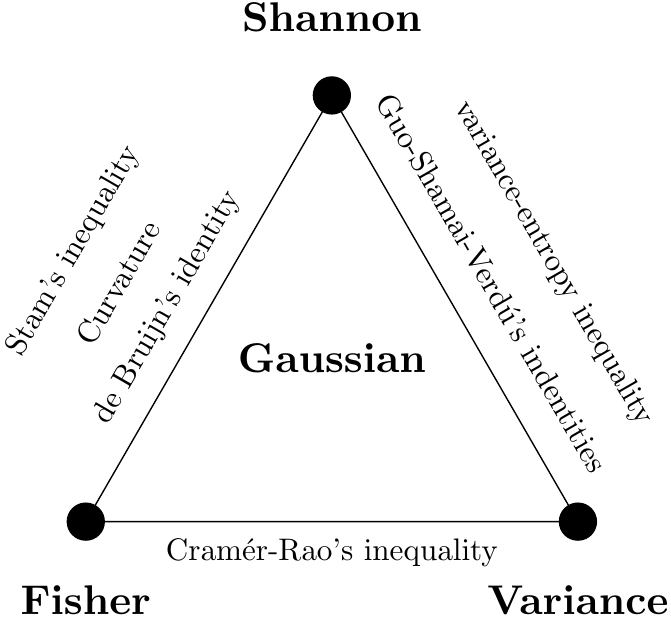}}
\caption{Classical  ``informational triangle''  that  schematically depicts  the
  hugely  used  information  measures  (at  the  vertices),  and  the  classical
  inequalities and  identities that  links these measures  (at the  edges).  The
  Gaussian  law is  central since  either the  identities concerns  the Gaussian
  channel, or the inequalities are saturated in the Gaussian context.}
\label{Triangle:fig}
\end{figure}

The study of  the notion of information related to a  random variable (r.v.), or
to  a  parameter attached  to  a  r.v., is  a  huge  long  outstanding field  of
investigation. The sense attributed to  ``information'' is closely linked to its
field  of  application.  The  most  usual  measures used  to  quantify  such  an
information can  be viewed  to be  the vertices of  a triangle,  as symbolically
depicted in figure~\ref{Triangle:fig}, and are
\begin{itemize}
\item The moments of a $d$-dimensional r.v. $X$, typically
\begin{equation}
\Esp[f(X)] = \int_{\Omega} f(x) \, p_X(x) \, dx
\end{equation}
for  some function  $f$ (independent  of the  pdf), where  $p_X$ stands  for the
probability  density function  (pdf) of  $X$  and $\Omega  \subset \Rset^d$  its
support; For $f(x) = x$, the mean  $m_X$ describes where the pdf is centered and
for $f(x) = \left( x-\Esp[X] \right) \left( x-\Esp[X] \right)^t$ where $\cdot^t$
stands  for  the  transposition\footnote{In   this  paper,  vectors  are  column
  vectors.}, the covariance matrix $C_X$ of  $X$ describes the spread of the pdf
around its mean: in some sense, these are two ``information measures'' regarding
the pdf. A typical associated measure of interest is the Mean-Square Error (MSE)
of an  estimator $\widehat{\theta}(X)$ of  a parameter $\theta$, built  using an
observed variable $X$ parametrized by $\theta$,
\begin{equation}
\MSE(\widehat{\theta}) = \Esp\left[ \left(\widehat{\theta} - \theta\right)
\left(\widehat{\theta} - \theta\right)^t \right]
\end{equation}
This quantity is widely used in estimation  in order to assess the quality of an
estimator for instance (its trace gives the ``power'' of the estimation error).
\item  The differential  Shannon entropy  of a  r.v. is  defined as~\cite{Sha48,
    CovTho06, Rio07}
\begin{equation}
H(X) = - \int_{\Omega} p_X(x) \, \log(p_X(x)) \, dx,
\end{equation}
and taking the exponential, one obtains the quantity known as the entropy power
\[
N(X)  = \frac{1}{2  \pi e}  \exp\left(\frac{2}{d} H(X)\right),
\]
which  is  generally  viewed  as  a  measure of  uncertainty.   Indeed  for  any
invertible (deterministic) matrix $A$ and any (deterministic) vector $b$ one has
$N(A X + b) = |A|^2 N(x)$  (where $|\cdot|$ stands for the absolute value of the
determinant).  Thus,  when $|A|$ goes to 0,  $AX + b$ tends  to be deterministic
and its uncertainty goes to 0. At the opposite, when $|A|$ goes to the infinity,
the law  of $X$ tends  to be  highly dispersed and  the uncertainty tends  to be
infinite.   $H$  can  also  be  viewed  as the  ``information''  brought  by  an
observation or outcome.   This quantity was naturally introduced  in the context
of communication, and the associate measure of particular interest is the mutual
information between  two random variables,  $\displaystyle I(X;Y) =  H(X)+H(Y) -
H(X,Y)$, \ie
\begin{equation}
I(X;Y) = \int_{\Omega} p_{X,Y}(x,y) \, \log\left( \frac{p_{X,Y}(x,y)}{p_X(x)
p_Y(y)} \right) \, dx \, dy
\end{equation}
This measure is fundamental as it quantifies the information transmitted through
a  communication channel while  the maximal  input-output information  gives the
channel   capacity.  The  mutual   information  can   be  written   through  the
Kullback-Leibler divergence, also called relative entropy~\cite{CovTho06},
\begin{equation}
\Dkl(p\|q) =  \int_{\Omega} p(x) \, \log\left( \frac{p(x)}{q(x)} \right) \, dx
\end{equation}
that is a kind  of distance of a pdf $p$ to a pdf  $q$ that serves as reference:
$\Dkl(p\|q) \ge 0$ with  equality if and only if $p =  q$ almost everywhere, but
it    is    not    symmetric    and    does    not    satisfy    the    triangle
inequality~\cite{CovTho06}.
\item   The    last   ``vertex''    of   the   informational    triangle   given
  figure~\ref{Triangle:fig}  is the  Fisher information  matrix relatively  to a
  $n$-dimensional parameter $\theta$ attached  to a r.v. $X$~\cite{Sta59, Kay93,
    Fri04},
\begin{equation}
J_\theta(X) = \int_\Omega \Big[ \nabla_\theta \log(p_X(x)) \Big] \,
\Big[\nabla_\theta \log(p_X(x)) \Big]^t \, p_X(x) \, dx
\end{equation}
where  $\nabla_\theta  f =  \left[  \frac{\partial  f}{\partial \theta_1}  \quad
  \ldots  \quad  \frac{\partial   f}{\partial  \theta_n}\right]^t$  denotes  the
gradient  of $f$  versus $\theta  = [\theta_1  \quad \ldots  \quad \theta_n]^t$.
Function  $\nabla_\theta  \log(p_X)$ is  known  as  the  score function  (versus
$\theta$) of the pdf.  This matrix  is highly popular in the estimation field as
it quantifies  the information on $\theta$ carried  by the r.v. $X$.  As we will
see in  a few  lines, it  allows to bound  the variance  of an  estimator.  When
$\theta$ is  a location parameter  (for example the  mean of the  variable), the
gradient in $\theta$  can be replaced by  a gradient in $x$, the  Fisher is then
known as the nonparametric Fisher information matrix, simply denoted by $J(X)$.
\end{itemize}

Although they come from different scientific fields (probability theory, digital
communications,  estimation,\ldots) these  quantities are  generally  related to
each  other, very  often by  inequalities,  as symbolically  represented by  the
``edges''  of the  triangle in  figure~\ref{Triangle:fig}.  Among  the classical
ones, given  in~\cite{DemCov91, CovTho06} for  instance, let us mention  some of
them :
\begin{itemize}
\item The moment-entropy relations $N(X)  \le |C_X|$ where $|\cdot|$ denotes the
  determinant. This relation is also detailed and extended in a series of papers
  by  Lutwak   et  al.~\cite{LutYan04,   LutYan05,  LutYan07,  LutLv12}   or  by
  Bercher~\cite{Ber12:06_1,Ber13}.
%
\item The Cram\'er-Rao inequality that links the variance of a r.v. --- or of an
  estimator  --- to the  Fisher information,  $C_X -  J(X)^{-1} \ge  0$ \  and \
  $\MSE(\widehat{\theta}) -  J_\theta(X)^{-1} \ge 0$ (in  the unbiased context),
  where  $A  \ge  0$  means  that matrix  $A$  is  positive~\cite{Kay93}.   This
  inequality   also    gave   rise   to    extensions~\cite{LutYan05,   LutLv12,
    Ber12:06_1,Ber12:06_2,Ber13}.
\item The Fisher information appears to be the curvature of the Kullback-Leibler
  divergence:  for  a pdf  parametrized  by $\theta  \in  \Theta$,  for a  given
  $\theta_0  \in  \Theta$,  the  second-order  Taylor  series  expansion  versus
  $\theta$ in  $\theta = \theta_0$ writes  $\Dkl ( p_\theta \|  p_{\theta_0} ) =
  \frac12 ( \theta - \theta_0 )^t J_{\theta_0}(X) ( \theta - \theta_0 ) + o ( \|
  \theta - \theta_0 \|^2 )$~\cite{CovTho06, Rio07}.
\item The Stam's inequalities lower  bound the product between the entropy power
  and the trace or the determinant of the Fisher information matrix~\cite{Sta59,
    Dem90, DemCov91,  CovTho06}, $N(X) \Tr(J(X))  \ge d$ where $\Tr$  stands for
  the trace operator and $N(X) |J(X)|^{\frac{1}{d}} \ge 1$.  As for the previous
  inequality,   the  Stam's   one   were   also  extended   by   Lutwak  or   by
  Bercher~\cite{LutYan05, LutLv12, Ber12:06_1, Ber13}.
\item The two following relations we are precisely interested in here, due to de
  Bruijn  and Guo  et  al.  respectively,  are  remarkable since  they link  two
  information measures  by identities rather than inequalities.   They deal with
  the Gaussian  channel, as depicted  in figure~\ref{GaussianChannel:fig}, where
  $G$ is a zero-mean standard Gaussian noise independent of the input $X$. Under
  some regularity assumptions,  the de Bruijn's identity links  the variation of
  the entropy  of the output's pdf with  respect to the noise  variance, and its
  Fisher  information~\cite{Sta59}.  The  Guo-Shamai-Verd\'u relation  links the
  variations of  the input-output mutual  information with respect to  the input
  power and the MMSE of the estimation of $X$ from the output $Y$, $\MMSE(X|Y) =
  \MSE(E[X|Y])$  (see~\cite{Kay93}).   For the  Gaussian  scalar context,  these
  relations are recalled in figure~\ref{GaussianChannel:fig}.
\begin{figure}[htbp]
\centerline{\includegraphics[width=6cm]{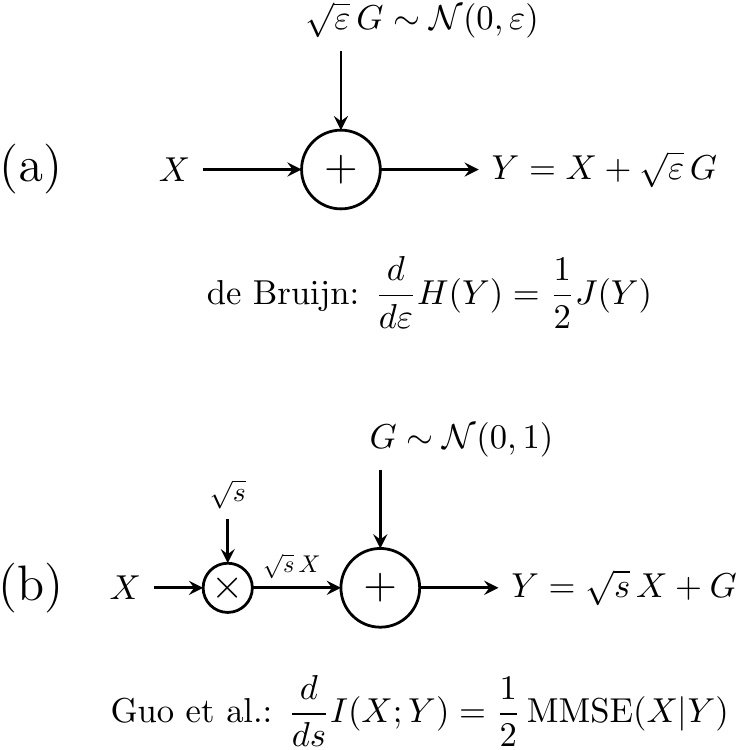}}
\caption{The Gaussian  channel, where the input  $X$ is corrupted  by a Gaussian
  noise $G$. (a):  In the de Bruijn's approach, the variation  of the entropy is
  characterized  versus the  noise variance  $\varepsilon$.  (b):  In  the Guo's
  approach, the noise variance is  fixed and the pre-amplification $\sqrt{s}$ of
  the input can vary: the  variations of the mutual information is characterized
  versus the Minimal Mean-Square Error of the estimation of $X$ using $Y$.}
\label{GaussianChannel:fig}
\end{figure}
Several alternative formulations exists  in terms of Kullback-Leibler divergence
versus  Fisher  divergence~\cite{Bar86,   Joh04}.\newline  These  relations  are
precisely at the heart of our paper.  Our goal is to generalize them outside the
usual ``Gauss-Shannon-Fisher'' context.
\end{itemize}
In  these  relationships,  the  Gaussian  play  a central  role  since  all  the
above-mentioned inequalities are saturated  for Gaussian random variables, while
the identities concern the Gaussian channel.

The  de Bruijn's identity  is very  important as,  for instance,  it was  in the
elements  involved in  the proof  of the  entropy  power inequality~\cite{Bla65,
  Sta59, DemCov91,  CovTho06}, and  in the proof  of the  above-mentioned Stam's
inequality as well~\cite{Bla65, DemCov91, CovTho06}.  All these inequalities can
also serve  as a  basis to prove  the central limit  theorem~\cite{Bar84, Bar86,
  JohBar04, Joh04}.

Because the de  Bruijn's identity or its Guo's  version expresses the variations
of  the  output  entropy  of   the  Gaussian  channel  (or  mutual  input-output
information),  it  finds natural  applications  in  communications.  Indeed,  as
stressed in~\cite{PalVer06}  and the series of  papers by the same  team, the de
Bruijn identity thus  allows to assess the behavior of  a canal versus variation
of the noise  amplitude, and thus its robustness faced  to noise. The divergence
version is  also used  to assess  the behavior of  such a  channel subject  to a
mismatch  between  an assumed  input  and  a  true one~\cite{Guo09,Ver10}.  This
identity and some possible extensions showed also its importance through various
applications, as for instance given by Park et al.~\cite{ParSer12:06, ParSer12},
Brown  et al.  in~\cite{BroDas06}  or Guo  et al.   in~\cite{GuoSha05:09}, among
others.   We can  mention for  instance,  the derivation  of Cram\'er-Rao  lower
bounds from a Bayesian perspective (BCRLB)  or from a frequentist point of view,
min-max optimal training sequences for channel estimation and synchronization in
the presence  of unknown noise distribution, applications  for turbo (iterative)
decoding schemes, generalized  EXIT charts and power allocation  in systems with
parallel non-Gaussian noise channels, application in graph theory.

\

While  Shannon entropy is  widely used  in communication,  there is  currently a
re-emergence of the use of  more general entropic tools, in particular R\'enyi's
and    Havrda-Charv\`at-Dar\'oczy-Tsallis's   entropies~\cite{HavCha67,   Dar70,
  Tsa88}. These generalized entropies  find applications in various domains such
as  in  statistical   physics~\cite{Tsa88,  Tsa99,  MarPla00,  PorPla96,  Jiz03,
  JizAri04},    in    multifractal    analysis~\cite{Har01}   or    in    signal
processing~\cite{CovTho06,  Cam65, Ber09}.   As the  Kullback-Leibler divergence
quantifies  the  ``distance'' between  a  pdf  relatively  to another  known  as
reference, other  divergences can also  quantify such a distance,  in particular
that  of the  class of  Csisz\'ar (or  Ali-Silvey)~\cite{Csi67,  AliSil66} given
later on in definition~\ref{PhiDivergence:def} and denoted $\Df$.  As previously
mentioned,   the  generalization   of   such  entropies,   together  with   some
generalizations of  the moments,  gave rise later  on to generalizations  of the
moment-entropy   inequalities~\cite{LutYan04,   LutYan05,   LutYan07,   LutLv12,
  Ber12:06_1,Ber13}.

To  generalize  the  Fisher information,  one  can  imagine  to start  from  the
definition~\ref{PhiDivergence:def}, eq.~\eqref{PhiDivergence:eq}  given later on
of  the  $\phi$-divergences and  to  make a  second  order  Taylor expansion  of
$\Df(p_\theta   \|  p_{\theta_0})$   in   $\theta  =   \theta_0$   as  for   the
Kullback-Leibler divergence.  However,  for Csisz\'ar's divergences sufficiently
smooth, the curvature coincides  again with the Fisher information~\cite{Vaj73},
showing the strength of this last quantity.  This direction is thus not relevant
to  generalized   the  Fisher  information.   Nevertheless,  in   spite  of  the
fundamental character  of this  measure, following pioneer  works from  Boeke or
Vajda~\cite{Boe77, Vaj73},  generalizations of  the Fisher information  began to
appear.   These  extensions  were  construct intrinsically  from  the  R\'enyi's
entropies and then used to extend information-theoretic results on the ``edges''
of the  ``informational triangle''  of figure~\ref{Triangle:fig}, such  that the
Cram\'er-Rao inequality~\cite{LutYan05, LutLv12, Ber12:06_1,Ber12:06_2,Ber13} or
the Stam's inequality~\cite{LutYan05, LutLv12, Ber12:06_1, Ber13} in the R\'enyi
context.  Although not  presented as a generalization of  the Fisher divergence,
one  can  find precisely  a  quantity in~\cite{Guo09}  that  appears  as such  a
generalization.  It  came from of a  possible generalization of  the de Bruijn's
identity  in  the scalar  context.   We  will see  later  on  that our  proposed
generalizations of this identity in terms of divergence makes in fact appear the
expression  of~\cite[th.~15]{Guo09}. Both  generalizations of  the informational
measures gave  rise to generalizations of  their links, or were  built to obtain
such generalizations.

\

Although many  parts of the informational  triangle of figure~\ref{Triangle:fig}
were generalized,  as far as we know,  a few generalizations of  the de Bruijn's
identity were  proposed.  In~\cite{Guo09}, Guo  proposed a version  by extending
their previous  version in terms  of mutual information  and MMSE in  the scalar
context, through Csisz\'ar's divergences. A generalization of the Shannon mutual
information--MMSE version in  the non-Gaussian context was also  proposed by the
same author~\cite{GuoSha05:09}. Finally, one can mention a generalization of the
identity  for  a  law   satisfying  a  nonlinear  heat  equation~\cite{JohVig07,
  Ber13:08}.   But  in this  non  linear  context,  connecting the  extended  de
Bruijn's identity to a noisy communication channel fails.

In our work,  we are interested in answering the  following questions.  (i) What
happens  in terms  of  robustness of  the  Gaussian channel  if  we use  general
divergences (or relative entropies) to  characterize the system?  (ii) Are there
equivalent results for  more general channels rather than  the Gaussian channel?
The main result  of the paper is  that the de Bruijn's identity  extends both to
general divergences  rather than the  Kullback-Leibler one, and to  more general
channels rather  than the Gaussian  one.  In these cases,  particular quantities
appear which we name  as $\phi$-Fisher information and $\phi$-Fisher divergences
and we will show that these  extensions contain special cases, such as the usual
Fisher information, the $\alpha$-Fisher gain~\cite{Ham78}, or a recently defined
Jensen-Fisher divergence~\cite{SanZar12}.  As the R\'enyi's entropies showed its
importance   in  various  field   of  applications   in  particular   in  signal
processing~\cite{CovTho06, Cam65,  Ber09}, extending  the de Bruijn  identity in
such  a context,  and  far beyond  this  last one,  open  perspectives in  these
applications in the light of the proposed extensions.

The  known  results and  the  extensions proposed  here  are  summarized in  the
following table.
\begin{table}[h]
\begin{center}
\begin{tabular}
{
|>{\Centering}m{.1\textwidth}
||>{\Centering}m{.183\textwidth}
|>{\Centering}m{.18\textwidth}|
}
\hline
&
Shannon \linebreak Fisher & $\phi$-entropies \linebreak $\phi$-Fisher\\
\hline\hline
Gaussian\linebreak channel &
Stam~\cite{Sta59}\linebreak
Barron (scalar)~\cite{Bar86}\linebreak
Johnson (scalar)~\cite{Joh04} &
Guo (scalar)~\cite{Guo09}\newline Sec.~3\\
\hline
Cauchy\linebreak channel &
Johnson (scalar)~\cite{Joh04} & Sec.~3 \&~4\\
\hline
L\'evy \linebreak channel &
Johnson (scalar)~\cite{Joh04} &  Sec.~3 (scalar)\\
\hline
2$^{\mbox{\small nd}}$ order PDE channels & Sec.~3 \&~4 & Sec.~3 \&~4\\
\hline
\end{tabular}
\end{center}
\end{table}

The paper is organized  as follows.  In section~\ref{def_nota:sec}, the notation
and assumptions  used throught  the paper  are shown. Then,  we will  recall the
definition  of the  $\phi$-entropies, $\phi$-divergences  with  their associated
$\phi$-Fisher  informations  and  $\phi$-Fisher divergences,  respectively.   In
section~\ref{scalar:sec}  we will  reformulate  the usual  de Bruijn's  relation
related to the scalar Gaussian channel~\cite{Sta59} in terms of the more general
$\phi$-divergences      due       to      Csisz\'ar~\cite{Csi67}      or      to
Ali-Silvey~\cite{AliSil66}  (see also~\cite{AmaNag00,  LieVaj06}).  In  the same
section  we  will go  beyond  the Gaussian  noisy  channel  and consider  noises
characterized by a  more general pdf.  We show two  instances of the generalized
version  extending   relations  proposed  by   Johnson  for  Cauchy   or  L\'evy
channels~\cite{Joh04}.   In section~\ref{multivariate:sec},  we will  go  a step
further, proposing multivariate extensions in which both the spatial coordinates
and   the   noise    parameter   are   vectors.    We   will    then   show   in
section~\ref{multivariate:sec},  that this  generalization encompasses  both the
multivariate de  Bruijn's identity~\cite{Sta59, CovTho06,  Rio07, DemCov91}, the
Guo's  one~\cite{GuoSha05},  as   well  as  other  extensions  due   to  Guo  et
al.~\cite{GuoSha05},  Palomar \&  Verd\'u  ~\cite{PalVer06} or  Johnson~\cite[\S
5.3]{Joh04}.


\section{Definitions and notations}
\label{def_nota:sec}

\subsection{Notations and assumptions}
\label{notat:subsec}

Throughout the  paper use the  following notations and assumptions  (except when
specified or when additional assumptions are required):
\begin{itemize}
\item The  probability laws are assumed to  admit a density with  respect to the
  Lebesgue measure.
\item The  probability density function (pdf)  is denoted $p$  when dealing with
  entropies, and $p_1$ and $p_0$ --the reference-- when dealing with divergences
  and  are defined  over sets  $\Omega \subseteq  \Rset^d$,  $\Omega_0 \subseteq
  \Rset^d$  and   $\Omega_1  \subseteq  \Rset^d$  respectively,   where  $d  \in
  \Nset^*$(multivariate context).
\item The pdfs are supposed to be parametrized by a (common) vectorial parameter
  $\theta \in \Theta \subseteq \Rset^n$, $n\in \Nset^*$.
\item  The ``states''  spaces $\Omega$,  $\Omega_0$ and  $\Omega_1$  are assumed
  to be independent of $\theta$.
\item  We assume  that  $\Omega_0 \subset  \Omega_1$,  that is  $p_0(x)  = 0  \:
  \Rightarrow  \: p_1(x)  = 0$  (the probability  measure attached  to  $p_1$ is
  absolutely continuous with respect to that attached to $p_0$).
\item Densities $p$ and $p_0$ are  assumed to vanish in the boundary of $\Omega$
  and $\Omega_0$, respectively.
\item When necessary, densities $p$,  $p_0$ and $p_1$ are assumed differentiable
  or twice differentiable with respect to $\theta$ and/or with respect to $x$.
\item The notation $\cdot^t$ denotes  the transposition operation of a vector or
  a matrix, $\Tr$ is the trace operator and $|\cdot|$ denotes the absolute value
  of the determinant of a matrix.
\item The gradient or jacobian vs  $\theta$ of a function $f: \Omega \rightarrow
  \Rset^k$   is   defined   as   $\nabla_\theta  f   =   \left[   \frac{\partial
      f_j}{\partial\theta_i} \right]_{i,j}$ so  that for $n=1$, $\nabla_\theta f
  = \frac{\partial  f^t}{\partial\theta}$.  The gradient  or jacobian vs  $x$ is
  defined similarly via the partial derivative vs $x$.
\item The Hessian  matrix vs $\theta$ of function  $f: \Omega \rightarrow \Rset$
  is           defined          as          $\Hess_\theta           f          =
  \left[\frac{\partial^{2}f}{\partial\theta_i  \partial\theta_j} \right]_{i,j}$,
  so that  for $n=1$, $\Hess_{\theta}  f=\frac{\partial^2 f}{\partial\theta^2}$.
  The  Hessian  vs  $x$  is  defined  similarly via  the  second  order  partial
  derivative vs $x$.
\item  The logarithm  function will  be denoted  $\log$, without  specifying its
  base; the  choice has no importance,  provided the same one  is considered for
  all the quantities that interplay.
\item The  entropic functional $\phi: [0  ; +\infty) \rightarrow  \Rset$ we will
  introduce  in  few  lines  needs  to  be  convex.   In  the  whole  paper,  we
  additionally assume  that it is of  class $C^2$, so that  the convexity writes
  $\phi''  \ge  0$ ($\cdot'$  and  $\cdot''$ denote  the  first  and the  second
  derivative, respectively).
\end{itemize}


\subsection{Definitions}
\label{def:subsec}

To  extend  the  de   Bruijn's  identity  to  generalized  $\phi$-entropies  and
$\phi$-divergences,  we  need  first  to  introduce these  quantities,  and  the
extensions of  the Fisher information as  well. What we  will call $\phi$-Fisher
information and  $\phi$-Fisher divergences are quantities  that appear naturally
when   $\phi$-entropies/divergences  are   used  instead   of  the   Shannon  or
Kulback-Leibler   divergence   to   characterize   the  channels   depicted   in
figure~\ref{GaussianChannel:fig} (and extensions in the non-Gaussian context).

\

Let  us  start   with  the  definition  of  the   $\phi$-entropies  and  of  the
$\phi$-divergences   of  the  the   Csisz\'ar's  class~\cite{Csi67}   (see  also
Salicr\'u~\cite{Sal87}):

\begin{definition}[$\phi$-entropies] Here, we assume additionally\footnote{ This
    condition is necessary (but not sufficient) to insure the convergence of the
    integral.}  that $\phi(0)  = 0$. The $\phi$-entropies of a  pdf $p$ are then
  defined as
\label{Phientropies:def}
\begin{equation}
\Hf(p) = - \int_\Omega \phi(p(x))\, dx.
\label{Phientropies:eq}
\end{equation}
where $\phi$ is the so-called entropic functional.
\end{definition}

Famous particular cases of  $\phi$-entropies are, among many others~\cite{Bas89,
  Bas13, Sal87}:
\begin{itemize}
\item The Shannon entropy~\cite{Sha48}, given by \ $\phi(l) = l\log (l)$;
\item   The  Havrda-Charv\'at~\cite{HavCha67},   or   Dar\'oczy~\cite{Dar70}  or
  Tsallis~\cite{Tsa99}  entropies,  denoted in  the  sequel HCDT\footnote{It  is
    worth to point out  that the R\'enyi entropies, $\displaystyle R_{\alpha}(p)
    = \frac{1}{1-\alpha} \log \left(  \int_\Omega p^\alpha(x) \, dx \right)$, is
    closely  connected to  the HCDT  entropies $T_\alpha$  since  $R_\alpha(p) =
    \frac{1}{1-\alpha}\log\left(  1  -  (1-\alpha) T_{\alpha}(p)  \right)$.}   \
  obtained for  $\displaystyle \phi(l) =  \frac{l^\alpha-l}{\alpha-1},\ \alpha >
  0$  (one can  even consider  the  situation $\alpha  \le 0$  when $\Omega$  is
  bounded);
\item The Kaniadakis entropies~\cite{Kan01}, given by \ $\displaystyle \phi(l) =
  \frac{l^{1+\alpha}-l^{1-\alpha}}{2 \alpha},\ -1<\alpha<1$.
\end{itemize}

\begin{definition}[$\phi$-divergences                    (Csisz\'ar~\cite{Csi67},
  Ali-Silvey~\cite{AliSil66})]
\label{PhiDivergence:def}
The  $\phi$-divergences   between  two  pdfs   $p_1$  and  $p_0$,   or  relative
$\phi$-entropies,  relatively to  pdf $p_0$,  are defined  as\footnote{One often
  finds  a more  general  definition  under the  form  $\displaystyle h\left(  -
    \int_\Omega  \phi\left(  \frac{p_1(x)}{p_0(x)}   \right)  \,  p_0(x)  \,  dx
  \right)$  where $h$  is an  increasing  function.  We  restrict here  to $h  =
  \operatorname{Id}$ the  identity, so that some usual  $\phi$-divergences are a
  monotonous  function  of  the   divergences  defined  here.   Note  also  that
  in~\cite{Csi67}, in the scalar context,  the integration is over $\Rset$ using
  the  convention \  $0 \,  \phi(0/0) =  0$; moreover  to avoid  the restriction
  $\Omega_1 \subseteq \Omega_0$,  Csisz\`ar also imposes the convention  \ $0 \,
  \phi(a/0) = a \lim_{u \to +\infty} \phi(u)/u$~\cite{Csi67}.}
\begin{equation}
\Df(\lrd) = \int_{\Omega_0} \phi\left( \frac{p_1(x)}{p_0(x)} \right) p_0(x) \, dx
\label{PhiDivergence:eq}
\end{equation}
\end{definition}

Well-known  cases  of such  divergences  are  the following~\cite{Bas89,  Bas13,
  Sal87, KumChi05}:
\begin{itemize}
\item The Kullback-Leibler divergence~\cite{CovTho06, Csi67} given by $\phi(l) =
  l \log(l)$;
\item  The exponential  of  the  so-called R\'enyi's  divergences  and a  linear
  function   of   the  Hellinger's   divergences   (or   simply  the   Hellinger
  integral)~\cite{CovTho06,  Ren61, Csi67, LieVaj06}  for $\phi(l)  = l^\alpha$,
  $\alpha    >    1$   (see    also    Tsallis~\cite{Tsa99}    or   Havrda    \&
  Charv\'at~\cite{HavCha67});
\item  The  Jensen-Shannon   divergence~\cite{CovTho06,  Csi67,  LieVaj06},  for
  $\phi(l) = \frac{l}{2} \log l - \frac{l+1}{2} \log \frac{l+1}{2}$ ,
\item   Vajda   divergences~\cite{Vaj73,   LieVaj06},   given  by   $\phi(l)   =
  |l-1|^\alpha$, $\alpha  \ge 1$ (including  the total variation  divergence for
  $\alpha=1$, and the Pearson divergence for $\alpha=2$).
\end{itemize}
Such  divergences  have  many   common  properties,  and  among  them,  assuming
additionally\footnote{ This is  not a restriction since for  any convex function
  $\widetilde{\phi}$    defined   on    $\Rset_+^*$,    function   $\phi(x)    =
  \widetilde{\phi}(x) - \widetilde{\phi}(1) x$ remains convex and $D_\phi(p_1 \|
  p_0)  =  D_{\widetilde{\phi}}(p_1  \|  p_0)  -  \widetilde{\phi}(1)$  is  only
  affected by  a shift.}  that  $\phi(1) = 0$,  from the Jensen  inequality such
$\phi$-divergences  are  nonnegative,  and zero  if  and  only  if $p_1  =  p_0$
(a.e.)~\cite{Csi67}.   We  let  the  reader  to  references~\cite{Bas89,  Bas13,
  LieVaj06} for  a brief  panorama and for  some applications of  divergences in
signal processing, physics and statistics.

\

Let  us  now turn  to  the generalization  of  the  second information  quantity
appearing in the de Bruijn's identity, namely the Fisher information.

\begin{definition}[$\phi$-Fisher information matrix]
\label{ParametricPhiFisherInformation:def}
We define  the $\phi$-Fisher  information matrix  of a pdf  $p$ relatively  to a
parameter $\theta$ by
\begin{equation}
\Jf_\theta(p) = \int_\Omega \Big[ \nabla_\theta \log p(x) \Big] \Big[
\nabla_\theta \log p(x) \Big]^t \, \big[ p(x) \big]^2\, \phi''(p(x)) \, dx
\label{ParametricPhiFisherInformation:eq}
\end{equation}
\end{definition}

As an illustration  we now show some particular cases that  already exist in the
literature.
\begin{itemize}
\item  Obviously, in the  Shannon context  $\phi(p) =  p \log  p$, so  that $p^2
  \phi''(p) = p$: one recovers the usual Fisher information matrix $J$.
\item  In  the  context of  the  HCDT  entropies,  $\phi(p) =  \frac{p^\alpha  -
    p}{\alpha-1}$ and thus $p^2 \phi''(p) = \alpha \, p^\alpha$. It appears that
  the          $\phi$-Fisher           information          matrices          of
  definition~\ref{ParametricPhiFisherInformation:def}    coincide    with    the
  $q$-Fisher   information   matrices   proposed   recently   by   Johnson   and
  Vignat~\cite[def.~3.2]{JohVig07} (where their $q$ and our $\alpha$ are related
  by $\alpha  = 2  q - 1$  and up  to a normalization  coefficient) or  with the
  $(2,\lambda)$-Fisher   information   matrices    introduced   by   Lutwak   et
  al.~\cite[eqs.~(13)-(18)]{LutLv12} (where their $\lambda$ and our $\alpha$ are
  related  by  $\alpha  =  2  \lambda-1$  and  up  to  a  factor  $\alpha$;  see
  also~\cite[eq.~(7)]{LutYan05} in the scalar context).
\end{itemize}

\begin{definition}[$\phi$-Fisher divergence matrices]
\label{ParametricPhiFisherDivergence:def}
We  define the  $\phi$-Fisher divergence  matrices  between two  pdfs $p_1$  and
$p_0$, relatively to parameter $\theta$ and the reference pdf $p_0$ by
\begin{equation}
\Jf_\theta(\lrd) = \int_{\Omega_0} \left[ \nabla_\theta \log \left(
\frac{p_1(x)}{p_0(x)} \right) \right] \left[ \nabla_\theta \log \left(
\frac{p_1(x)}{p_0(x)} \right) \right]^t \left[ \frac{p_1(x)}{p_0(x)} \right]^2
\, \phi''\left( \frac{p_1(x)}{p_0(x)} \right) \, p_0(x) \, dx
\label{ParametricPhiFisherDivergence:eq}
\end{equation}
\end{definition}
When $\theta$  is a location  parameter $\nabla_\theta \equiv \nabla_x$  and the
$\phi$-Fisher  information   and  Fisher  divergence  matrices   reduce  to  the
corresponding   nonparametric   ones,    denoted   $\Jf(p)$   and   $\Jf(\lrd)$,
respectively.

Some  particular  cases  of  $\phi$-Fisher  divergences  were  proposed  in  the
literature, in specific contexts, as follows:
\begin{itemize}
\item  For the  entropic  function $\phi$  of  the Kullback-Leibler  divergence,
  $(p_1/p_0)^2 \phi''(p_1/p_0) p_0 = p_1 $ and thus the $\phi$-Fisher divergence
  that  corresponds   to  the   same  function  $\phi$   is  the   usual  Fisher
  divergence~\cite{CovTho06, Joh04, Rio07}.
\item  In  the R\'enyi  context,  with the  R\'enyi  index  $\alpha$ (or  HCDT),
  $(p_1/p_0)^2  \phi''(p_1/p_0)  p_0  \propto  p_1^\alpha p_0^{1-\alpha}$  is  a
  geometric mean  of densities $p_1$ and  $p_0$, leading, up  to a normalization
  factor, to the $\alpha-$Fisher gain introduced by Hammad in~\cite{Ham78}.
\item   Note  finally   that  for   the  Jensen-Shannon   context,  $(p_1/p_0)^2
  \phi''(p_1/p_0)  p_0  \propto \frac{p_1  p_0}{p_1+p_0}$  is  an harmonic  mean
  leading to  a very recently defined  Jensen-Fisher divergence $J^{(JS)}(\lrd)$
  by     S\'anchez-Moreno    et     al.~\cite{SanZar12}.     In~\cite{SanZar12},
  $J^{(JS)}(\lrd)$  was  introduced  by  pure analogy  with  the  Jensen-Shannon
  divergence under  the form  $$ J^{(JS)}(\lrd) =  \frac12 J \left(  p_0 \left\|
      \frac{p_0  + p_1}{2}  \right.   \right)  + \frac12  J  \left( p_1  \left\|
      \frac{p_0 + p_1}{2} \right. \right) =  \frac12 J(p_0) + \frac12 J(p_1) - J
  \left( \frac{p_0  + p_1}{2}  \right)$$ and was  used for  physical description
  purposes.
\end{itemize}

Both these  matrices are symmetric positive  definite and vanish if  and only if
$p_1=p_0$  a.e.  Moreover,  as for  the usual  Fisher divergence,  one  can also
define the (scalar) $\phi$-Fisher divergences  as the trace of the $\phi$-Fisher
divergence  matrices.   Thus,   obviously,  the  $\phi$-Fisher  divergences  are
nonnegative.

Note that,  as shown  in~\cite[eq. (26)]{Ham78} for  the usual  Fisher matrices,
both  the  $\phi$-divergences  and  the $\phi$-Fisher  divergence  matrices  are
invariant by the same biunivocal transformation of both $p_1$ and $p_0$.

\

As  already evoked  in the  introduction, a  generalization of  the  de Bruijn's
identity  in  the  scalar   context  for  the  $\phi$-divergences  and  Gaussian
channel\footnote{More  precisely,  a  general   channel  is  considered,  as  in
  figure~\ref{GaussianChannel:fig}-(a),  where  $\varepsilon  \to 0$.   In  this
  limit, the  result lies on the heat  equation followed by the  output pdf, and
  the channel can  be viewed as approximately Gaussian (in  the second order and
  provided  the  noise   has  a  finite  variance).}   has   been  made  by  Guo
in~\cite{Guo09} where,  although no notion  of Fisher information  is explicitly
mentioned, the  derivative of the  $\phi$-divergences with respect to  the noise
parameter is linked with  the nonparametric $\phi$-Fisher information, this last
quantity being expressed  in terms of the difference  of score functions $\nabla
\log(p)$.


\section{Extension of  the scalar de  Bruijn's identity to  $\phi$-entropies and
  $\phi$-Fisher informations}
\label{scalar:sec}

In this section, we  focus on the scalar context for both  the state $x$ and the
parameter  $\theta$, \ie  $d=n=1$.  This  restriction allows  to increment  by a
first  step the  de Bruijn's  identity, while  the general  case  (including the
scalar one) will  be the object of  the next section.  In this  section, we will
assume  that  the  quantities  that interplay  (entropies,  divergences,  Fisher
informations) exist. This assumption requires  conditions on the pdfs and on the
entropic functional $\phi$ that cannot be  given in a general setting; they must
be studied case by case.

\

Let  us consider  firstly the  Gaussian  channel as  in the  de Bruijn's  primal
version, as  done by Guo's in  some sense in  its extension~\cite{Guo09}, before
generalizing the result for a class of more general noises.


\subsection{Gaussian noise}
\label{gauss_noise:subsec}
The key point of  the de Bruijn's identity for the Gaussian  channel is that the
pdf $p$ of the output follows the heat equation
\begin{equation}
\frac{\partial p}{\partial \theta} = \frac12 \frac{\partial^2 p}{\partial x^2}.
\label{heat_equation:eq}
\end{equation}
Reproducing the  same steps than for  the usual de Bruijn  identity, writing the
$\phi$-entropies of $p$, performing the  derivative of this quantity once versus
the parameter $\theta$,  and using the heat equation,  one obtains the following
extension that we name $\phi$-de Bruijn's identity,
\begin{proposition}[$\phi$-de Bruijn's identity]
\label{PhiDeBruijnGauss:prop}
Consider a  pdf $p$ satisfying the  heat equation~\eqref{heat_equation:eq}, such
that  $\frac{\partial}{\partial\theta}  \phi(p)$  is $\theta$-locally  uniformly
integrable\footnote{By  this terminology,  we express  that for  any  compact $K
  \subset \Theta$,  this partial  derivative is integrable  vs $x$  on $\Omega$,
  uniformly  vs $\theta  \in K$.   This  allows to  interchange integration  and
  derivation  vs  $\theta$~\cite[\S~63]{Woo34}.   In  practice,  the  sufficient
  condition that $\left|  \frac{\partial}{\partial\theta} \phi(p) \right| \le g$
  for any  $\theta \in  K$ with  $g$ integrable and  independent of  $\theta$ is
  often used, invoking thus the  dominated convergence theorem together with the
  mean    value    theorem.},    and    such    that    both    $\phi(p)$    and
$\frac{\partial}{\partial x}  \phi(p)$ vanish in the boundary  of $\Omega$. Then
its $\phi$-entropy and $\phi$-Fisher information fulfill
\begin{equation}
\frac{d}{d\theta} \Hf(p) = \frac12 \Jf(p). 
\label{DeBruijnGauss:eq}
\end{equation}
\end{proposition}
\begin{IEEEproof}
  This  case   is  a  particular   case  of  proposition~\ref{PhiDeBruijn:prop},
  section~\ref{multivariate:sec}, proved in appendix~\ref{PhiDeBuijn:app}.
\end{IEEEproof}

In   the   Shannon   context,    one   recovers   the   original   de   Bruijn's
identity~\cite{Sta59}.

Note that  from the assumptions  that $p$ vanishes  in the boundary  of $\Omega$
together with $\phi(0) = 0$, the vanishing assumption of $\phi(p)$ is indeed not
a strong  restriction.  The other assumptions  have to be studied  case by case,
given the explicit form of $p$ and $\phi$.

A particular situation  of the general one depicted  in this proposition, widely
used in communication theory, occurs  when considering the output for a Gaussian
noisy channel, $Y =  X + \sqrt{\theta} G$ since the pdf  of the output satisfies
the        heat       equation~\cite{Bar84,        Bar86}        (see       also
appendix~\ref{HeatEquation:app}).   Clearly,  in  such  a case,  the  regularity
conditions  stated in the  propositions for  pdf $p_Y$  imply conditions  on the
input pdf,  depending on the entropic  functional $\phi$.  For  instance, in the
Shannon case there were shown to be  true by Barron, provided that the input has
a finite  variance~\cite[Lemma~6.3]{Bar84}.  In the general  $\phi$ context, the
steps  of  Barron  are  more  difficult  to apply.   However,  it  is  shown  in
appendix~\ref{BoundaryConditions:app} that $\phi(p_Y)$  vanishes in the boundary
of the domain. Moreover,  assuming that there exists some $k \in  (0 \; 1)$ such
that  $u^k  \phi'(u)  \to  0$  when   $u  \to  0$,  the  vanishing  property  of
$\frac{\partial}{\partial  y}\phi(p_Y)$ in  the  boundary is  also insured  (see
appendix~\ref{BoundaryConditions:app}).  This  last condition on  $\phi'$ is not
very  restrictive, applying  for the  entropies frequently  used, such  that the
Shannon entropy,  the HCDT entropy (provided  $k > 1-\alpha$)  or the Kaniadakis
entropy (provided that $k > \kappa$), in others.

\

As done  for the  Kullback-Leibler divergence in~\cite{Joh04},  this proposition
can be recast in terms of $\phi$-divergences as follows.
\begin{proposition}[$\phi$-de Bruijn's identity in terms of divergences]
\label{PhiDivDeBruijnGauss:prop}
Let $p_0$ and $p_1$ parametrized by the same parameter $\theta$, both satisfying
the heat  equation~\eqref{heat_equation:eq}, such that $\frac{\partial}{\partial
  \theta} \left[ p_0 \phi\left(\lr\right) \right]$ is $\theta$-locally uniformly
integrable, and  such that both $p_0 \phi\left(\lr\right)$  and $\nabla_x \left[
  p_0 \phi\left(\lr\right)  \right]$ vanish in  the boundary of  $\Omega$. Then,
their $\phi$-divergences and $\phi$-Fisher divergences satisfy
\begin{equation}
\frac{d}{d \theta} \, \Df(\lrd) = - \, \frac12 \Jf(\lrd).
\label{DeBruijn_Gauss_Df_partiel:eq}
\end{equation}
\end{proposition}
\begin{IEEEproof}
  This  case  is  a  particular case  of  proposition~\ref{PhiDivDeBruijn:prop},
  section~\ref{multivariate:sec}, proved in appendix~\ref{PhiDivDeBuijn:app}.
\end{IEEEproof}
Again,  a particular  situation  arises in  the  context of  the Gaussian  noisy
channel.  As previously mentioned the pdf of the output of this Gaussian channel
satisfies the heat  equation.  For instance, in a  mismatch context, considering
that $X_0$ is the  assumed input of the channel, while the  true input is $X_1$,
and  noting   $p_0$  and  $p_1$  the   pdfs  of  the   respective  outputs,  the
$\phi$-divergence measures  a kind  of distance between  the assumed  output pdf
$p_0$  (that  serves as  the  reference)  and the  true  one  $p_1$. Hence,  the
$\phi$-Fisher  information gives  the variation  of this  mismatch  measure with
respect  to the  noise amplitude.   Since  $J^{(\phi)} \ge  0$, the  proposition
states that the consequence of the mismatch decreases with $\theta$, the rate of
decreasing being precisely  given by this $\phi$-Fisher information.   If $X_0 =
0$,  the divergence  measures the  decrease  of the  distance to  a Gaussian  as
$\theta$ increases,  which is  a key point  used in  some proofs of  the central
limit  theorem when  dealing with  the  Kullbach-Leibler divergence~\cite{Bar86,
  Joh04, JohBar04}.   It has also been shown  that in the limit  $\theta \to 0$,
the proposition  apply for non Gaussian  noises with finite variance  and in the
small amplitude  noise limit  $\theta \to 0$  since in  this limit the  pdf also
satisfy the  heat equation (and thus  for the output  pdf as well)~\cite{Guo09}.
As shown in this last reference,  the $\phi$-Fisher information can be viewed as
a  mean-square  distance  between  the   outputs'  pdfs,  but  averaged  over  a
``deformed'' distribution  instead of the  reference one.  Finally, anew  in the
Shannon context, one  recovers the original de Bruijn's  identity formulation in
terms of divergences of~\cite{Bar86, Joh04} (there, the reference pdf, $p_0$, is
a Gaussian  of variance $\theta$ and $p_1$  as the output pdf).   Let us finally
mention that  in the case of  Jensen-Shannon divergence and in  the scalar case,
eq.~\eqref{PhiDivDeBruijnGaussMulti:eq}  reduces to  the  Sanchez-Moreno et  al.
version of such de Bruijn's identity~\cite[eq. (7)]{SanZar12}.

Note that once  again, the conditions of the proposition are  to be studied case
by case according  to the considered entropic functional $\phi$  and the pdfs of
the inputs $X_0$ and $X_1$ as well.


\subsection{Beyond  the  Gaussian  noise:   extension  to  more  general  scalar
  non-Gaussian channels}

Here,         we         extend         propositions~\ref{PhiDeBruijnGauss:prop}
and~\ref{PhiDivDeBruijnGauss:prop} to a more general set up in which the channel
noise is non-Gaussian.   Indeed, although the most common noise  in nature is of
Gaussian  type, there  exists  others whose  probability  distribution does  not
follow  the heat  equation  but  still have  associated  a partial  differential
equation  (PDE)  which,  in turn,  is  the  clincher  to obtain  de  Bruijn-type
identities. We will consider the general case as well as two particular cases of
non-Gaussian noises, L\'evy  and Cauchy, whose corresponding PDE  have a similar
structure to that of the heat equation~\cite{Joh04}.

\

For both versions of  the de Bruijn's identity, the key point  is that $p_0$ and
$p_1$ follow the same second order PDE given by
\begin{equation}
\alpha_1(\theta) \, \frac{\partial}{\partial \theta} \, p(x) + \alpha_2(\theta)
\, \frac{\partial^2}{\partial\theta^2} \, p(x) = \frac{\partial}{\partial x}
\left( \beta_1(x,\theta) \, p(x) \right) + \beta_2(\theta) \,
\frac{\partial^2}{\partial x^2} \, p(x)
\label{pde_pdf:eq}
\end{equation}
Note    that   the   PDE~\eqref{pde_pdf:eq}    reduces   to    a   Fokker-Planck
equation~\cite{Ris89}, when $\alpha_2 = 0$  and $\alpha_1 = 1$, where $-\beta_1$
is  the   drift  and   where  $2   \,  \beta_2$  is   the  diffusion,   that  is
state-independent  in this  case,  the  heat equation  being  a particular  case
($\beta_1 = 0$ and with $\beta_2$ being constant).

\

Now, propositions~\ref{PhiDeBruijnGauss:prop} and~\ref{PhiDivDeBruijnGauss:prop}
can be generalized one step further as follows:
\begin{proposition}[Generalized scalar $\phi$-de Bruijn identity]
\label{PhiDeBruijnScalar:prop}
Let a pdf $p$ satisfying the PDE~\eqref{pde_pdf:eq} where the drift $\beta_1$ is
state-independent  ($\beta_1(x,\theta)  =   \beta_1(\theta)$),  such  that  both
$\frac{\partial}{\partial\theta}                   \phi(p)$                  and
$\frac{\partial^2}{\partial\theta^2}  \phi(p)$  are  $\theta$-locally  uniformly
integrable,  and  such  that  both $\phi(p)$  and  $\frac{\partial}{\partial  x}
\phi(p)$ vanish  in the  boundary of $\Omega$.   Then, the  $\phi$-entropies and
$\phi$-Fisher information of pdf $p$ satisfy the identity
\begin{equation}
\alpha_1(\theta) \, \frac{d}{d\theta} \Hf(p) \, + \, \alpha_2(\theta) \,
\frac{d^2}{d\theta^2} \, \Hf(p) = \beta_2(\theta) \, \Jf(p) \, - \,
\alpha_2(\theta) \, \Jf_\theta(p).
\label{Scalar_DeBruijn_General_Fentropies:eq}
\end{equation}
\end{proposition}
\begin{IEEEproof}
  This    is  a  particular   case  of  proposition~\ref{PhiDeBruijn:prop},
  section~\ref{multivariate:sec}, proved in appendix~\ref{PhiDeBuijn:app}.
\end{IEEEproof}
When   $\alpha_2   =   0$,   the   $\theta$-local   uniform   integrability   of
$\frac{\partial^2}{\partial\theta^2} \phi(p)$ is  unnecessary and, similarly, if
$\beta_2 = 0$, no condition on $\frac{\partial}{\partial x} \phi(p)$ is required
(see the proof of the proposition).

Condition $\phi(p) \to 0$ in the  boundary of $\Omega$ is not restrictive due to
the assumption $\phi(0) = 0$ and the vanishing assumption of $p$ in the boundary
of $\Omega$.   The other regularity  conditions stated in the  proposition imply
conditions on the pdf $p$, depending on the entropic functional $\phi$, and must
be  studied  case   by  case.  To  this  end,  one  can   follow  the  steps  of
Barron~\cite{Bar84,    Bar86}     recalled    and    slightly     extended    in
appendix~\ref{BoundaryConditions:app} as a guidance.

\

As in the  heat equation context, this proposition can again  be recast in terms
of divergences as follows:
\begin{proposition}[Generalized  scalar $\phi$-de  Bruijn identity  in  terms of
  divergences]
\label{PhiDivDeBruijnScalar:prop}
Let $p_0$ and $p_1$ two pdfs,  with the same parameter $\theta$, both satisfying
PDE~\eqref{pde_pdf:eq}  and   such  that  both  $\frac{\partial}{\partial\theta}
\left[          p_0          \phi\left(\lr\right)          \right]$          and
$\frac{\partial^2}{\partial\theta^2}  \left[  p_0 \phi\left(\lr\right)  \right]$
are  $\theta$-locally uniformly  integrable,  and such  that  both $\beta_1  p_0
\phi\left(\lr\right)$    and    $\frac{\partial}{\partial    x}    \left[    p_0
  \phi\left(\lr\right) \right]$ vanishes in  the boundary of $\Omega$. Then, the
$\phi$-divergences and  $\phi$-Fisher divergences of  pdf $p_1$ with  respect to
$p_0$ fulfill the relation
\begin{equation}
\alpha_1(\theta) \, \frac{d}{d\theta} \Df(\lrd) + \alpha_2(\theta) \,
\frac{d^2}{d\theta^2} \Df(\lrd) = \alpha_2(\theta) \, \Jf_\theta(\lrd) -
\beta_2(\theta) \, \Jf(\lrd)
\label{Scalar_DeBruijn_General_Fdivergence:eq}
\end{equation}
\end{proposition}
\begin{IEEEproof}
  This case is again a particular case of proposition~\ref{PhiDivDeBruijn:prop},
  section~\ref{multivariate:sec}, proved in appendix~\ref{PhiDivDeBuijn:app}.
\end{IEEEproof}
As  for the  entropic  version of  the  proposition, when  $\alpha_2  = 0$,  the
$\theta$-local uniform  integrability of $\frac{\partial}{\partial\theta} \left[
  p_0 \phi\left(\lr\right) \right]$ is unnecessary, and similarly, if $\beta_2 =
0$, no condition on $\frac{\partial}{\partial x} \left[ p_0 \phi\left(\lr\right)
\right]$ is required (see the proof of the proposition).

Note that here  again, the conditions of the proposition are  to be studied case
by case according  to the considered entropic functional $\phi$  and the pdfs of
the inputs $p_0$ and $p_1$ as well.

\

It is first interesting to note that these identities apply again in the context
of  a noisy channel  as in  figure~\ref{GaussianChannel:fig}-(a), but  where the
noise pdf  satisfies PDE~\eqref{pde_pdf:eq} in the  context of state-independent
$\beta_1$. Indeed, in this case, writing the output pdf as a convolution between
the   input  and   noise  pdfs,   and  provided   that  the   last   is  regular
enough\footnote{The first and  second derivative vs $\theta$ and  vs $x$ must be
  $\theta$-locally  uniformly integrable  and  $x$-locally uniformly  integrable
  respectively.  Noting  that the  output pdf can  be obtained as  a convolution
  between the pdf of the input and  that of the noise, a sufficient condition is
  that  the  partial and  second  order partial  derivatives  of  the noise  are
  (locally)  uniformly  bounded.   Thus,  the  integrand  are  dominated  by  an
  integrable  function  proportional  to  the  input  pdf,  with  a  coefficient
  independent of the parameter ($\theta$ or  $x$).} one can show that the output
pdf also  satisfy PDE~\eqref{pde_pdf:eq}  (see the steps  given in  the Gaussian
case,  appendix~\ref{HeatEquation:app}).  Thus,  these identities  apply  to the
output of such a  general channel, or to the output pdf  relatively to the noise
pdf, provided the  conditions required by the propositions  are satisfied (these
ones impose conditions on the input that can only be studied when the pdf of the
noise is explicitly known). In other words, these results include and generalize
the   standard    de   Bruijn's    identity~\cite{Bar84}   as   well    as   the
Guo-Shamai-Verd\'u's  relations~\cite{GuoSha05,  Guo09}  either to  non-Gaussian
noise, or  to $\phi$-entropies and divergences,  or both. They  also include and
generalize identities  derived by Johnson in~\cite{Joh04} for  Cauchy and L\'evy
channels in  the context of Kullback-Leibler  divergence as we will  show in few
lines. The first version of  the Guo-Shamai-Verd\'u's identity is also recovered
and  extended  by  considering  the  output  pdf  and  the  pdf  of  the  output
conditionaly      to       the      input      in       the      context      of
figure~\ref{GaussianChannel:fig}-(b).


\paragraph{L\'evy channel}
Consider again  the channel of the form  fig.~\ref{GaussianChannel:fig}, but now
subject to L\'evy noise with  scale parameter $\theta^2$, \ie $\theta^2 L$ where
$L$  is  a  standard L\'evy  r.v..   $\theta^2  L$  has  then  the pdf  $p(x)  =
\frac{\theta  \,  \exp \left(  -  \frac{\theta^2}{2  x}\right)}{\sqrt{2 \pi}  \,
  x^{\frac32}}$ defined  on $\Rset_+$~\cite{SamTaq94}.  One can  easily see that
both  this pdf,  and  more specially,  the pdf  of  the output  $Y$ satisfy  the
parabolic  differential equation~\cite{Joh04} (see  also the  steps used  in the
Gaussian case recalled appendix~\ref{HeatEquation:app})
\begin{equation}
\frac{\partial^2}{\partial \theta^2} \, p(x) = 2 \, \frac{\partial}{\partial x}
p(x)
\label{parabolic_equation:eq}
\end{equation}
As an immediate  consequence of proposition~\ref{PhiDivDeBruijnScalar:prop}, the
$\phi$-entropy of the output and its $\phi$-Fisher information are linked by the
relation
\begin{equation}
\frac{d^2}{d\theta^2} \, \Hf(p)  = \Jf_\theta(p)
\label{Scalar_DeBruijn_Levy_Phientropies:eq}
\end{equation}
Similarly,  for two  output pdfs  $p_0$  and $p_1$  of the  L\'evy channel  (for
instance  when  the  input  is  respectively of  L\'evy  and  arbitrary),  their
$\phi$-divergences and $\phi$-Fisher divergences satisfy the relation
\begin{equation}
\frac{d^2}{d\theta^2} \, \Df(\lrd)  = \Jf_\theta(\lrd)
\label{Scalar_DeBruijn_Levy_PhiDivergence:eq}
\end{equation}
The  identity  directly  links  the  curvature of  the  $\phi$-entropies  (resp.
$\phi$-divergences)  with the  $\phi$-Fisher  information (resp.   $\phi$-Fisher
divergences).  For a  L\'evy distributed input vs an arbitrary  input and in the
context               of               Kullback-Leibler              divergence,
relation~\eqref{Scalar_DeBruijn_Levy_PhiDivergence:eq}    is    precisely   that
obtained  by Johnson  in~\cite[Th.  5.5]{Joh04}.  Again,  to study  some of  the
conditions required by  the proposition, in the entropy  context, one can follow
the sketch appendix~\ref{BoundaryConditions:app}.


\paragraph{Cauchy channel}
Consider again  the channel  fig.~\ref{GaussianChannel:fig}, but now  subject to
Cauchy noise  with scale parameter  $\theta$, \ie $\theta  \, C$ where $C$  is a
standard  Cauchy r.v..   $\theta \,  C$ has  the pdf  $p(x)  = \frac{\theta}{\pi
  \left( \theta^2  + x^2  \right)}$~\cite{SamTaq94} so that  both this  pdf and,
specially, the  pdf of the  output, satisfy the Laplace  (elliptic differential)
equation~\cite{Joh04} (the  very same steps  used in the Gaussian  case recalled
appendix~\ref{HeatEquation:app} allows to this conclusion)
\begin{equation}
\frac{\partial^2}{\partial \theta^2} \, p(x) = - \frac{\partial^2}{\partial x^2}
p(x).
\label{laplace_equation:eq}
\end{equation}
Thus,   as  a   consequence  of   proposition~\ref{PhiDeBruijnScalar:prop},  the
$\phi$-entropy and  $\phi$-Fisher information  of the output  are linked  by the
relation
\begin{equation}
\frac{d^2}{d \theta^2} \, \Hf(p)  = \Jf(p) + \Jf_\theta(p)
\label{Scalar_DeBruijn_Cauchy_Phientropies:eq}
\end{equation}
Similarly  for two  output  pdfs $p_0$  and  $p_1$ of  the  Cauchy channel  (for
instance when the input is respectively of Cauchy and arbitrary),
\begin{equation}
\frac{d^2}{d \theta^2} \, \Df(\lrd)  = \Jf(\lrd) + \Jf_\theta(\lrd)
\label{Scalar_DeBruijn_Cauchy_PhiDivergence:eq}
\end{equation}
Note  again  that,  now,  the  identity  directly links  the  curvature  of  the
$\phi$-entropies  (resp.   divergences)  with  the  sum of  the  parametric  and
nonparametric $\phi$-Fisher informations  (resp.  divergences).  Here again, for
a      Cauchy     distributed     input      vs     an      arbitrary     input,
relation~\eqref{Scalar_DeBruijn_Cauchy_PhiDivergence:eq}    reduces    to   that
obtained  by   Johnson  in~\cite[Th.   5.6]{Joh04}.    Dealing  with  entropies,
following  the  very same  steps  that in  appendix~\ref{BoundaryConditions:app}
allows to conclude that the  boundary conditions required by the proposition are
satisfied.


\section{From the scalar case to the multidimensional context}
\label{multivariate:sec}

In this section, we generalize  the previous results to the general multivariate
context,  both for the  state $x$  ($d \ge  1$) and  parameter $\theta$  ($n \ge
1$).  To this  aim, as  for the  previous section,  the approach  relies  on pdf
satisfying    a    second    order    PDE    with    the    same    form    than
eq.~\eqref{pde_pdf:eq}.  But since the  gradient operators  lead to  vectors (or
matrices in the context of Jacobian  matrices) and the Hessian operators lead to
matrices, one  have to introduce operators  in order to sum  quantities with the
same dimension.

More  precisely,  we  consider  pdf  $p$,  parametrized  by  a  vector  $\theta$
satisfying the following PDE,
\begin{equation}
\opL_1 \big( \nabla_\theta p(x) \big) + \opL_2 \big( \Hess_\theta p(x) \big) =
\opK_1 \big( \nabla_x \big[ \beta_1(x,\theta) p(x) \big] \big) + \opK_2 \big(
\Hess_x p(x) \big)
\label{pdegen_pdf:eq}
\end{equation}
where $\opL_i$ and $\opK_i$ are  linear operators acting on vectors or matrices,
dependent on $\theta$ or not but independent on the state $x$,
\begin{itemize}
\item $\opL_1: \Rset^n \longrightarrow \Rset^k$ \ and \ $\opL_2: \Rset^{n \times
    n} \longrightarrow \Rset^k$,
\item $\beta_1: \Rset^d \times \Rset^n \longrightarrow \Rset^l$,
\item  $\opK_1: \Rset^{d  \times l}  \longrightarrow \Rset^k$  \ and  \ $\opK_2:
  \Rset^{d \times d} \longrightarrow \Rset^k$
\end{itemize}

for some  $l \in \Nset$  and $k \in  \Nset$. For instance, an  operator $\opK_i$
and/or $\opL_i$  can be  the trace operator,  a right  and/or left product  by a
matrix  (possibly dependent  of $\theta$),  extraction of  a subvector  or  of a
submatrix, etc.

To get an  idea on pdfs satisfying a PDE  of the form eq.~\eqref{pdegen_pdf:eq},
consider  the  Gaussian  pdf  $$p(x)  =  \frac{1}{(2  \pi  \theta)^{\frac{d}{2}}
  |R|^{\frac12}}   \exp\left(-  \frac{1}{2  \theta}   x^t  R^{-1}   x  \right)$$
parametrized by the  scalar $\theta$.  Differentiating in $\theta$  on one hand,
differentiating twice  in $x$ on the  other hand, and using  the identity $\Tr(u
v^t) = u^t v$, one easily shows that $p$ satisfies the PDE
\begin{equation}
\nabla_\theta p = \Tr\left( R \, \Hess_x p \right)
\label{MultivariateHeatEquation:eq}
\end{equation}
Here, $\opL_2 = 0$, $\opL_1 = \opI$ is the identity, $\opK_1 = 0$ and $\opK_2(M)
= \Tr(R M)$ for any $M \in \Rset^{d \times d}$.

\

Note that,  in the particular context  of a state-independent  $\beta_1$, if the
input noise pdf of a  channel as in figure~\ref{GaussianChannel:fig} satisfies a
PDE of the  form eq.~\eqref{pdegen_pdf:eq}, the pdf of  the output satisfies the
same PDE\footnote{As  for the scalar case,  the gradient and Hessian  have to be
  $\theta$- and  $x$-locally uniformly integrable. Again, it  is sufficient that
  these quantities  are $\theta$ and $x$-locally uniformly  bounded.}.  Thus, in
this case we are in situation  to generalize the multivariate versions of the de
Bruijn identities related to a noisy communication channel.

\

If  $\theta$ is scalar,  PDE~\eqref{pdegen_pdf:eq} encompasses  the multivariate
Fokker-Planck  equation  with state-independent  diffusion  when  $\opL_1 =  0$,
$\opL_2 = \opI$, $-\beta_1:  \Rset^d \times \Rset \longrightarrow \Rset^d$ being
the drift, $\opK_1(M) = \Tr(M)$ for  any matrix $M \in \Rset^{d \times d}$, and,
denoting $\un^t  = [1 \quad \ldots  1]$, $\opK_2(M) = \frac12  \un^t D(\theta) M
\un$ for  any matrix  $M \in \Rset^{d  \times d}$  and for a  symmetric positive
definite  matrix  $D:  \Rset  \longrightarrow  \Rset^{d  \times  d}$  being  the
diffusion tensor~\cite{Ris89}.

\

Finally,   as    mentioned   in   the   previous    propositions,   the   scalar
PDE~\eqref{pde_pdf:eq}  is  a particular  example  where  $\opL_i$  ($i =  1,2$)
reduces  to   multiplication  by  $\alpha_i(\theta)$,  $\opK_2$   reduces  to  a
multiplication by $\beta_2(\theta)$ and $\opK_1 = \opI$.

\

In the multivariate context introduced here  above, we can now generalize the de
Bruijn's identities  in terms  of $\phi$-entropies and  $\phi$-divergences. From
the generalizations, we will then exhibits three particular examples, recovering
existing identities of the literature.


\subsection{Multivariate general de Bruijn's identities}
%
\begin{proposition}[Generalized multivariate $\phi$-de Bruijn's relation]
\label{PhiDeBruijn:prop}
Let  $p$ be  a pdf  that fulfills  the PDE~\eqref{pdegen_pdf:eq}  with $\beta_1$
state-independent  ($\beta_1(x,\theta)  =   \beta_1(\theta)$),  such  that  both
$\nabla_\theta   \phi(p)$  and   $\Hess_\theta  \phi(p)$   are  $\theta$-locally
uniformly integrable, and such that both $\phi(p)$ and $\nabla_x \phi(p)$ vanish
in  the  boundary of  $\Omega$.  Then,  its  $\phi$-entropies and  $\phi$-Fisher
information matrices satisfy the relation
\begin{equation}
\opL_1 \big( \nabla_\theta \Hf(p) \big) + \opL_2 \big( \Hess_\theta \Hf(p) \big)
= \opK_2 \left( \Jf(p) \right) - \opL_2 \left( \Jf_\theta(p) \right)
\label{PhiDeBruijn:eq}
\end{equation}
\end{proposition}
\begin{IEEEproof}
See Appendix~\ref{PhiDeBuijn:app}.
\end{IEEEproof}
Here again, when $\opL_2 =  0$ the local integrability of $\Hess_\theta \phi(p)$
is unnecessary and  similarly, if $\opK_2 = 0$,  the gradient $\nabla_x \phi(p)$
does  not need  to vanish  in the  boundary of  $\Omega$ (see  the proof  of the
proposition).

\

As  for  the previous  scalar  extensions of  the  de  Bruijn's identities,  the
proposition can be recast in terms of divergences as follows,
\begin{proposition}[Generalized  multivariate  $\phi$-de  Bruijn's relations  in
  terms of divergences]
\label{PhiDivDeBruijn:prop}
Let  two pdfs $p_1$  and $p_0$  be parametrized  by a  same vector  $\theta$ and
satisying  PDE~\eqref{pdegen_pdf:eq}, such that  both $\nabla_\theta  \left[ p_0
  \phi\left( \lr  \right) \right]$ and  $\Hess_\theta \left[ p_0  \phi\left( \lr
  \right)  \right]$ are  locally integrable,  and  such that  both $\beta_1  p_0
\phi\left( \lr \right)$ and $\nabla_x \left[ p_0 \phi\left( \lr \right) \right]$
vanish  in   the  boundary  of  $\Omega$.   Then,   the  $\phi$-divergences  and
$\phi$-Fisher divergence matrices satisfy the relation
\begin{equation}
\opL_1 \big( \nabla_\theta \Df(\lrd) \big) + \opL_2 \big( \Hess_\theta \Df(\lrd)
\big) = \opL_2 \left( \Jf_\theta(\lrd) \right) - \opK_2 \left( \Jf(\lrd) \right)
\label{PhiDivDeBruijn:eq}
\end{equation}
\end{proposition}
\begin{IEEEproof}
See Appendix~\ref{PhiDivDeBuijn:app}.
\end{IEEEproof}
Once again,  when $\opL_2 = 0$  the local integrability  of $\Hess_\theta \left[
  p_0 \phi\left(  \lr \right) \right]$  is unnecessary. Similarly, if  $\opK_2 =
0$, the gradient  $\nabla_x \left[ p_0 \phi\left( \lr  \right) \right]$ does not
need to vanish in the boundary of $\Omega$ (see proof).

\
	
It appears that~\eqref{PhiDivDeBruijn:eq} looks  somewhat similar to an extended
version of the de Bruijn's identity proposed by Johnson \& Vignat in the context
of R\'enyi's  entropies~\cite[eq. (11)]{JohVig07}. However  our extension cannot
recover  their  version since~\eqref{PhiDivDeBruijn:eq}  is  based on  densities
satisfying     the     second      order     linear     partial     differential
equation~\eqref{pde_pdf:eq},  while  the version  of~\cite{JohVig07}  lies on  a
nonlinear extension of the heat  equation (called $q$-heat equation, involving a
so-called  $q$-Fisher information) as  mentioned in  the introduction.   But, as
previously  mentioned,  the  extension  proposed  in~\cite{JohVig07}  cannot  be
related to channel as  in figure~\ref{GaussianChannel:fig} so easily. Indeed, if
the  noise  satisfied  the  nonlinear  differential equation  leading  to  their
extension, due to the nonlinear aspect, the output cannot satisfy this equation.

\

Note again that the conditions required  to apply the last two propositions have
to be studied  case by case, given  $p$, $p_0$, $p_1$ and $\phi$;  again, such a
study  can  be inspired  by  that  of  Barron~\cite{Bar84, Bar86}  recalled  and
slightly extended in appendix~\ref{BoundaryConditions:app}.


\subsection{Particular cases}


\subsubsection{Gaussian channel}
Let      us     consider      the     Gaussian      channel      depicted     in
figure~\ref{GaussianChannel:fig}-(a), where the  noise is now $\sqrt{\theta} G$,
with $G$  a Gaussian  vector with  zero-mean and covariance  matrix $R$.   It is
straightforward  to show  that the  both noise  pdf and  output pdf  $p$ follows
PDE~\eqref{pdegen_pdf:eq} with $\opL_1  = \opI$, $\opL_2 = 0$,  $\opK_1 = 0$ and
$\opK_2(M) = \frac12 \Tr(R \, M)$,  \ie the multidimensional version of the heat
equation  given eq.~\eqref{MultivariateHeatEquation:eq}.  Then, one  obtains the
extended vector version of the de Bruijn's identity:
\begin{equation}
\frac{\partial}{\partial \theta} \Hf(p) = \frac12 \Tr \left( R \Jf(p) \right)
\label{PhiDeBruijnGaussMulti:eq}
\end{equation}
In the  Shannon entropy context,  the usual versions~\cite{PalVer06,  Joh04} are
obviously  recovered, and  of course,  in the  scalar case  and for  the Shannon
entropy, the initial de Bruijn's  identity, as presented by Stam in~\cite{Sta59}
is naturally recovered.

Moreover,  the   divergence  version  of   de  Bruijn's  relation,   also  given
in~\cite{Joh04} in the scalar context and Shannon entropies, writes
\begin{equation}
\frac{\partial}{\partial \theta} \Df(\lrd) = - \frac12 \Tr \left( R \Jf(\lrd)
\right),
\label{PhiDivDeBruijnGaussMulti:eq}
\end{equation}
If $p_1$ is the pdf of the output of the Gaussian channel and $p_0$ the Gaussian
pdf with the same covariance than the noise, this result can be interpreted as a
convergence of the  output to the Gaussian as $\theta$  increases since $\Jf \ge
0$  and  $R  \ge 0$  implies  a  decrease  of  $\Df$. Hence,  the  $\phi$-Fisher
divergence  associated to  the  $\phi$-divergence (\ie  with  the same  entropic
functional $\phi$)  gives the speed of  convergence. For that reason,  it is not
surprising that  this relation was implied  (in the scalar  Shannon context), in
some  way, in  a proof  of the  central limit  theorem  (see~\cite{Joh04, Bar86,
  JohBar04, TulVer06, MadBar06} or references cited in).
	
Note  also that  another way  of thinking  consists in  considering  two similar
channels,  with respective  input $X_0$  and $X_1$,  and of  output respectively
$Y_0$ and  $Y_1$ with pdfs $p_0$ and  $p_1$.  Thus, instead of  working with the
output and the noise, one can  wish to compare the two different outputs leading
to a  tendency of convergence  of the two  outputs' pdfs as  $\theta$ increases,
with   a   convergence   rate   given   by   the   corresponding   $\phi$-Fisher
divergence.  Again, through  this point  of view,  the  $\phi$-Fisher divergence
allows  to assess  the behavior  of  the channel  versus a  mismatch between  an
assumed input and a true one.
	
As  far as  we know,  these  interpretations and  the consequences  in terms  of
central limit  theorem will let open  the question of the  interpretation of the
general       de       Bruijn's       relations~\eqref{PhiDeBruijnGaussMulti:eq}
and~\eqref{PhiDivDeBruijnGaussMulti:eq}.

As  mentioned when  dealing  with the  scalar  context, following  the steps  of
Barron~\cite{Bar84, Bar86}, we  show in appendix~\ref{HeatEquation:app} that the
pdf  of  the  output of  the  multivariate  Gaussian  channel also  satisfies  a
multivariate heat equation  and in appendix~\ref{BoundaryConditions:app} that in
this multivariate context the boundary  conditions are also satisfied (under the
weak assumption of the  existance of a $k \in (0 \;  1)$ such that $u^k \phi'(u)
\to 0$ when $u \to 0$, dealing with the second one).


\subsubsection{Cauchy  channel} Let us  consider the  Cauchy channel,  where the
channel noise is $\theta C$, where $C$ has the characteristic matrix $R$ and its
associated density is given by
\begin{equation*}
p(x) = \frac{\Gamma\left( \frac{d+1}{2} \right)}{\pi^{\frac{d+1}{2}}
|R|^{\frac12}} \frac{\theta}{(\theta^2 + x^t R^{-1} x)^{\frac{d+1}{2}}}
\end{equation*}
which  follows  PDE~\eqref{pdegen_pdf:eq}  with  $\opL_1=0$,  $\opL_2  =  \opI$,
$\opK_1 = 0$ and $\opK_2 = -\opI$, \ie
\begin{equation}
\frac{\partial^2}{\partial\theta^2} p(x) = - \Tr \left( R \, \Hess_x p(x) \right).
\label{pde_multi_cauchy:eq}
\end{equation}

Again, following  the very same  steps than that of  Barron~\cite{Bar84, Bar86},
recalled in appendix~\ref{HeatEquation:app}, allows to easilly show that the pdf
output of a  multivariate Cauchy channel satisfies the same  PDE than the Cauchy
noise.

Thus,  assuming that  the pdfs  satisfying  eq.~\eqref{pde_multi_cauchy:eq} also
satisfy  the  condition   required  by  proposition~\ref{PhiDeBruijn:prop},  the
$\phi$-informational quantities of these pdfs satisfy the relation
\begin{equation}
\frac{d^2}{d\theta^2 }\Hf(p) = \Tr \left( R \, \Jf(p) \right) + \Tr \left( R \,
\Jf_\theta(p) \right).
\end{equation}
Note that following the steps given in appendix~\ref{BoundaryConditions:app} for
the Gaussian channel, one can also  easilly show that in the multivariate Cauchy
context, the  boundary conditions of  the proposition are also  satisfied (under
the same assumption for $u^k \phi'(u)$).  Thus, for both the Cauchy pdf, or that
of the output of a Cauchy channel.

Similarly,     for      two     pdfs     $p_0$      and     $p_1$     satisfying
eq.~\eqref{pde_multi_cauchy:eq},  for instance  the  pdf of  the  output of  the
Cauchy channel and a Cauchy distribution with the same characteristic matrix, or
the  pdfs  of  two  different  outputs  (\eg  in  the  mismatch  context)  their
$\phi$-divergences and $\phi$-Fisher divergences satisfy
\begin{equation}
\frac{d^2}{d\theta^2} \Df(\lrd) = \Tr \left( R \, \Jf(\lrd) \right) + \Tr\left(
R \, \Jf_\theta(\lrd) \right).
\end{equation}

\subsection{Extended Guo-Shamai-Verdu's and extended Palomar-Verdu's relations.}

As we will see now, the Guo's relation of~\cite{Guo09} as well as the scalar and
vectorial variations  given in~\cite{GuoSha05, GuoSha05:09,  PalVer06, TulVer06}
are particular cases of proposition~\ref{PhiDivDeBruijn:prop}.

First  of all,  one can  notice that  when parameter  $\theta$ is  matricial for
instance, by a vectorization of this  matrix, such a case can be treated through
the formalism adopted  in this section.  Moreover, to  conserve the structure of
the  quantities, the vectors  or matrices  that appear  through the  gradient or
Hessian can be rearranged  in tensors (``de-vectorization'').  In this paragraph
we only  need to differentiate real-valued  function of matricial  argument $M =
\left[  m_{i,j} \right]_{i,j}$  for  which $\nabla_M  f$  is the  matrix of  the
partial   derivatives,   $\nabla_M    f=   \left[   \frac{\partial   f}{\partial
    m_{i,j}}\right]_{i,j}$. Thus, we do need to introduce a complicate tensorial
formalism.   Note in  particular that  if $f(M)  = g(x)$  with $x  = M  u$, from
$\frac{\partial  f}{\partial m_{i,j}} =  \sum_k \frac{\partial  g}{\partial x_k}
\frac{\partial x_k}{\partial m_{i,j}}$ we obtain $\nabla_M f = \left[ \nabla_x g
\right] u^t$.

Now,      we      consider      again      the     Gaussian      channel      of
figure~\ref{GaussianChannel:fig}-(b), but where the input $X$ is a random vector
and the multiplication  is matricial, of the  form $H B X$, and  where the noise
$N$ is Gaussian independent of $X$,  zero-mean and of covariance matrix $R$, \ie
the output  is $ Y  = H  B X +  N$.  In such  a communication model,  matrix $H$
represents  the  transmission  channel   (filtering,  etc.)   while  matrix  $B$
represents a pre-treatment  of the data before sending them  to the channel (\eg
beamforming), or can model the covariance matrix of the input (\eg that would be
$B B^t$). $H$ and $B$ are matrices that can be rectangular.

In~\cite{PalVer06},  the  authors are  interested  in  the relationship  between
$\nabla_H I(X ; Y)$  or $\nabla_B I(X ; Y)$ and the  MMSE matrix. Such relations
allows to study the robustness of the information transmission vs the channel or
vs the pre-treatment. These results can  be recovered and extended thanks to the
following proposition:
\begin{proposition}
\label{PhiGuo:prop}
Let  us  consider  a  multivariate  Gaussian  channel  of  the  same  form  than
figure~\ref{GaussianChannel:fig}-(b),  of   input  $X$   put  in  form   by  the
multiplication with a  matrix $\theta$, and corrupted by  an independent channel
noise $N$, zero-mean, of covariance matrix $R$ that does not depend of $\theta$,
\ie  $Y =  \theta  X +  N$.  Assume that  $\nabla_\theta  \left[ p_Y  \phi\left(
    \frac{p_{Y|X=x}}{p_Y}   \right)  \right]$   is   $\theta$-locally  uniformly
integrable,  that  both $y  p_Y  \phi\left(  \frac{p_{Y|X=x}}{p_Y} \right)$  and
$\nabla_y \left[ p_Y \phi\left( \frac{p_{Y|X=x}}{p_Y} \right) \right]$ vanish in
the boudary  of $\Omega$ and  that $\nabla_\theta \left[ \Df\left(  p_{Y|X=x} \|
    p_Y \right)  p_X \right]$  is $\theta$-locally uniformly  integrable.  Thus,
the generalized $\phi$-mutual input-output  information $\Df( p_{X,Y} \| p_X p_Y
)$ satisfies the relation
\begin{equation}
\left( \nabla_\theta \Df(p_{X,Y} \| p_X p_Y) \right) \theta^t = R^{-1} \theta
\MSE_\phi(X|Y) \theta^t
\label{PhiGuo:eq}
\end{equation}
with
\begin{equation}
\MSE_\phi(X|Y) = \int_{\Omega^2} \left( x - \Esp[X|Y=y] \right)^2 \left(
\frac{p_{X,Y}(x,y)}{p_X(x) p_Y(y)} \right)^2 \phi''\left(
\frac{p_{X,Y}(x,y)}{p_X(x) p_Y(y)} \right) \, p_X(x) \, p_Y(y) \, dx \, dy
\label{MSEPhi:eq}
\end{equation}
\end{proposition}
Again, $\MSE_\phi(X|Y)$  can be interpreted as  a generalized $\phi$-mean-square
error matrix, as defined in~\cite{PalVer06} for the classical case.
\begin{IEEEproof}
  The proof is detailed in  appendix~\ref{PhiGuo:app}.  It lies on the fact that
  the   conditional    pdfs   $p_{Y|X=x}$    and   $p_Y$   satisfy    the   same
  PDE~\eqref{pdegen_pdf:eq}. The  result is thus almost a  direct consequence of
  proposition~~\ref{PhiDivDeBruijn:prop}.
\end{IEEEproof}

Now, for  $\theta = H  B$ together  with the fact  that for any  scalar function
$g(H)$ it holds that $\nabla_H g = \nabla_\theta g B^t$, we derive the relation
\begin{equation}
\nabla_H \Df\left( p_{X,Y} \| p_X p_Y\right) \, H^t = R^{-1} H B \MSE_\phi(X|Y) B^t H^t
\label{VerduGeneH:eq}
\end{equation}
that is nothing but~\cite[eq.  (21)]{PalVer06} up to the right multiplication by
$H^t$. Similarly, from $\nabla_B g = H^t \nabla_\theta g$ we obtain
\begin{equation}
\nabla_B \Df\left( p_{X,Y} \| p_X p_Y\right) \, B^t H^t = H^t R^{-1} H B
\MSE_\phi(X|Y) B^t H^t
\label{VerduGeneB:eq}
\end{equation}
that is nothing but~\cite[eq.  (22)]{PalVer06} up to the right multiplication by
$H^t$.

It  is  left as  a  future  investigation the  study  of  the simplification  of
Eq.~\eqref{PhiGuo:eq}  in the multivariate  case since  in this  case it  is not
feasible to simply eliminate $\theta^t$  from both sides of this equation, given
that  $\theta$  is a  matrix.   Just  in the  case  in  which  $\theta$ is  tall
(including square) and  has full rank, it is possible to  multiply both sides of
Eq.~\eqref{PhiGuo:eq} by  $\theta$ and then  (as the resulting square  matrix is
invertible)  to simplify  each side  by $(\theta^t  \theta)^{-1}$.  This  is for
instance always true in the scalar case.

Note that in the  scalar context of figure~\ref{GaussianChannel:fig}, the result
of Guo~\cite[Th.~3]{Guo09} is thus recovered in the Shannon case and extented to
$\phi$-divergences  noting  that  for  $\theta  =  \sqrt{s}$  and  noting  that,
$\frac{\partial}{\partial  s} = \frac{\partial}{\partial  \theta} \frac{\partial
  \theta}{\partial s} = \frac{1}{2 \sqrt{s}} \frac{\partial}{\partial\theta}$.



\section{Conclusions}

In this paper we have  proposed multidimensional generalizations of the standard
de Bruijn's identity obtained via the so-called $\phi$-entropies and divergences
of Csisz\'ar (also Salicr\'u) class, within a scalar and vectorial framework. We
first showed  that, in the scalar case  and for the Gaussian  noisy channel, the
derivative of  the $\phi$-entropy (divergence)  can be written as  a generalized
version of the  Fisher information (divergence) and that  these relations can be
considered as  a natural extension of  the classical de  Bruijn's relation where
both the noise and output pdfs  follow the heat equation. Then, we have proposed
a  further step  by considering  non-Gaussian noises  of pdfs  governed  by more
general linear second-order  PDE than the heat equation, both  in the scalar and
multivariate context  (for both the state  and the parameter).  We thus obtained
extended  versions of  the de  Bruijn's identity  as well  as extensions  of the
Guo-Shamai-Verd\'u's  relation  that  link  the gradient  of  the  $\phi$-mutual
information to the generalized $\phi$-mean-square error.

The physical interpretation of the  extended de Bruijn's identities remains open
as  well as  their  potential implications  and  applications. Nevertheless,  we
believe  that the  extensions shown  can broaden  the perspective  on  the usual
applications   in  statistics,  estimation,   communication  theory   or  signal
processing in a wider sense~\cite{CovTho06, Cam65, Ber09, ParSer12:06, ParSer12,
  BroDas06, GuoSha05:09, Guo09, Ver10, JohBar04, PalVer06}.


\section*{Acknowledgements}

I. V.  Toranzo acknowledges the support of the Spanish Ministerio de Educaci\'on
under the program FPU 2014 and the BioTic International Mobility Program for PhD
students. This work has also  been partially supported by the LabEx PERSYVAL-Lab
(ANR-11-LABX-0025-01) funded by the French program Investissement d'avenir.


\appendices


\section{Proof of the generalized multivariate $\phi$-de Bruijn's identities.}

In the proof of the propositions, we will very often use the divergence theorem,
under various forms. In order to have it in mind, we recall it here:

\begin{lemma*}[Divergence theorem]
  Let $\Omega  \subset \Rset^d$ be a  region in space and  $\partial \Omega$ its
  boundary. Consider a vector  field $f: \Omega \longrightarrow \Rset^d$.  Then,
  the  volume  integral of  the  divergence,  $\div  f =  \sum_i  \frac{\partial
    f_i}{\partial x_i}$ over $\Omega$ is  related to the surface integral of $f$
  over the boundary $\partial\Omega$ (assumed piecewise smooth) through
\begin{equation}
\int_\Omega \div f \: d\omega = \int_{\partial\Omega} f^t n \: ds,
\label{DivergenceTheorem:eq}
\end{equation}
where  $f  =  [f_1 \quad  \cdots  \quad  f_d]^t$  and where  $n:  \partial\Omega
\longrightarrow \Rset^d$ is the normal vector to the surface $\partial\Omega$.

When  applied  to  $f  =  c   \,  \psi$  where  $c$  is  an  arbitrary  constant
$d$-dimensional vector and $\psi: \Omega \longrightarrow \Rset$, it leads to
\begin{equation}
\int_\Omega \nabla \psi \: d\omega = \int_{\partial\Omega} \psi \: n \: ds
\label{DivergenceTheoremNabla:eq}
\end{equation}

Finally, applying this last equation to any component of $\nabla \psi$ one obtains
\begin{equation}
\int_\Omega \Hess \psi \: d\omega = \int_{\partial\Omega} \nabla \psi \: n^t \: ds
\label{DivergenceTheoremHessian:eq}
\end{equation}
\end{lemma*}


\subsection{Formulation in terms of the $\phi$-entropies}
\label{PhiDeBuijn:app}

Remind  first  that here  $\beta_1(x,\theta)  =  \beta_1(\theta)$  so that  this
function,  being  independent on  $x$,  can  be  inserted in  operator  $\opK_1$
(dependent only on $\theta$). In other words, without loss of generality, we can
consider here that $\beta_1 = 1$.

\

We start from definition~\ref{Phientropies:def}, eq.~\eqref{Phientropies:eq}, of
the $\phi$-entropies and perform the derivative respect to parameter $\theta$,
\begin{equation*}
\nabla_\theta \Hf = - \nabla_\theta \int_\Omega \phi(p) \, dx = - \int_\Omega
\nabla_\theta \left[\phi(p)\right] \, dx
\end{equation*}
where the argument $x$ of the functions are omitted for readability purposes, as
well  as the  argument  of  the $\phi$-entropies.  The  interchange between  the
derivative   and  the   integral   follows  from   the  $\theta$-local   uniform
integrability assumption~\cite[\S~63]{Woo34}. This gives the expression
\begin{equation}
\nabla_\theta \Hf = - \int_\Omega \left( \nabla_\theta p \right) \, \phi'(p) \, dx
\label{d_dtheta_Phientropies:eq}
\end{equation}
Differentiating again  versus $\theta$,  using again the  $\theta$-local uniform
integrability assumption to differentiate under the integral, we obtain
\begin{equation*}
\Hess_\theta \Hf = - \int_\Omega \left[ \left( \Hess_\theta p \right) \,
\phi'(p) \: + \: \left( \nabla_\theta p \right) \left( \nabla_\theta p \right)^t
\, \phi''(p) \right] \, dx
\end{equation*}
that      is,      from     definition~\ref{ParametricPhiFisherInformation:def},
eq.~\eqref{ParametricPhiFisherInformation:eq}, of the $\phi$-Fisher information,
\begin{equation}
\Hess_\theta \Hf = - \int_\Omega \left( \Hess_\theta p \right) \, \phi'(p) \, dx
\: - \: \Jf_\theta
\label{d2_dtheta2_Phientropies:eq}
\end{equation}
($\Jf_\theta$  is  supposed to  exist,  thus  the integral  of  the  sum can  be
separated as the sum of the integrals).

Then,  we use successively  the linearity  of the  operators\footnote{The linear
  operators acting on matrices, elements of a finite dimensional Hilbert spaces,
  can be written  as finite linear combination of the  elements of the matrices,
  and thus, due to the finiteness, can be permuted with the integration (see for
  instance~\cite{HunNac01}).}  $\opL_i$, the PDE~\eqref{pdegen_pdf:eq} satisfied
by $p$,  relation $\left( \Hess_x p \right)  \phi'(p) = \Hess_x \phi(p)  \, - \,
\left( \nabla_x p \right) \left(  \nabla_x p \right)^t \phi''(p)$, the linearity
of            operators            $\opK_i$,            together            with
definition~\ref{ParametricPhiFisherInformation:def}    of    the   nonparametric
$\phi$-fisher information to obtain,
%
\begin{eqnarray*}
\opL_1\big( \nabla_\theta \Hf \big) + \opL_2\big( \Hess_\theta \Hf \big) & = & -
\int_\Omega \left[ \opL_1\big( \nabla_\theta p \big) + \opL_2\big( \Hess_\theta
p \big) \right] \phi'(p) \, dx - \opL_2\left( \Jf_\theta \right)\\[2.5mm]
& = & - \int_\Omega \left[ \opK_1\big( \nabla_x p \big) + \opK_2\big( \Hess_x p
\big) \right] \phi'(p) \, dx - \opL_2\left( \Jf_\theta \right)\\[2.5mm]
& = & - \int_\Omega \left[ \opK_1\big( \nabla_x \phi(p) \big) + \opK_2\big(
\Hess_x \phi(p) \big) \right] dx + \opK_2\left( \Jf \right) - \opL_2\left(
\Jf_\theta \right).
\end{eqnarray*}
Because   $\phi(p)$   vanishes  on   the   boundary   of   $\Omega$,  from   the
formulation~\eqref{DivergenceTheoremNabla:eq} of the divergence theorem, we have
$$ \int_\Omega \nabla_x \left[ \phi(p) \right] dx = 0$$ Thus, from linearity
of   $\opK_1$,   since   necessarily   $\opK_1(0)   =  0$,   we   can   conclude
that $$\int_\Omega  \opK_1\big( \nabla_x  \phi(p) \big) dx  = 0$$  Similarly, as
$\nabla_x  \left[ \phi(p)  \right]$ vanishes  on the  boundary of  $\Omega$, one
obtains  from formulation~\eqref{DivergenceTheoremHessian:eq} of  the divergence
theorem together  with the linearity of $\opK_2$  that $$\int_\Omega \opK_2\big(
\Hess_x \phi(p) \big) dx = 0$$ which finishes the proof.


\subsection{Formulation in terms of $\phi$-divergences}
\label{PhiDivDeBuijn:app}

First of all, let us mention  the following useful expression that we will often
utilize in the sequel to make shorter the algebra,
\begin{equation}
\nabla \left( \lr \right) = \frac{1}{p_0} \left( \nabla p_1 - \lr \, \nabla p_0
\right) = \lr \, \nabla \log\left( \lr \right)
\label{nablaL:eq}
\end{equation}
where the derivative $\nabla$ can be either vs $x$, or vs $\theta$.

\

We   start  now   the  proof   by  first   computing  the   derivative   of  the
$\phi$-divergences       given       in      definition~\ref{PhiDivergence:def},
eq.~\eqref{PhiDivergence:eq} with respect to $\theta$.  Using the $\theta$-local
uniform integrability assumption allowing to interchange the derivative with the
integral, we obtain the expression
\begin{equation}
\nabla_\theta \Df = \int_\Omega \left[ \left( \nabla_\theta p_0\right) \,
\phi\left(\lr\right) + \left( \nabla_\theta p_1 - \lr \, \nabla_\theta p_0
\right) \phi'\left(\lr\right) \right]\, dx.
\label{dD_dtheta:eq}
\end{equation}

Similarely,  expressing the  Hessian of  $p_0 \phi(p_1/p_0)$  starting  from its
gradient, using the $\theta$-local uniform integrability of the gradient of $p_0
\phi(p_1/p_0)$ to  interchange the differentiation in $\theta$  and the integral
in~\eqref{dD_dtheta:eq},   using  relation~\eqref{nablaL:eq}  to   simplify  the
notation     and     the     definition~\ref{ParametricPhiFisherDivergence:def},
eq.~\eqref{ParametricPhiFisherDivergence:def}, we obtain the Hessian
\begin{equation}
\Hess_\theta \Df = \int_\Omega \left[ \left( \Hess_\theta p_0\right)
\phi\left(\lr\right) + \left( \Hess_\theta p_1 - \lr \, \Hess_\theta p_0 \right)
\phi'\left(\lr\right) \right] dx + \Jf_\theta.
\label{d2D_dtheta2:eq}
\end{equation}

Then,  (i)  one  combines eqs.~\eqref{dD_dtheta:eq}  and~\eqref{d2D_dtheta2:eq},
(ii) one uses  the linearity of operators $\opL_i$ to  interchange them with the
integrations, (iii)  one uses PDE~\eqref{pdegen_pdf:eq} satisfied  by both $p_0$
and $p_1$, (iv) one observes that
\[
\nabla_x \left[  \beta_1 p_1  \right] \phi\left( \lr  \right) +  \left( \nabla_x
  \left[ \beta_1 p_1  \right] - \lr \nabla_x \left[  \beta_1 p_0 \right] \right)
\phi'\left(\lr\right) = \nabla_x \left[ \beta_1 p_0 \phi\left(\lr\right) \right]
\]
and (v) that,
\[
\Hess_x  p_0 \,  \phi\left(\lr\right)  +  \left(\Hess_x p_1  -  \lr \Hess_x  p_0
\right)  \phi'\left(\lr\right)  =  \Hess_x  \left[ p_0  \,  \phi\left(\lr\right)
\right]   -   \frac{p_1^2}{p_0}   \left[\nabla_x  \log\left(\lr\right)   \right]
\left[\nabla_x \log\left(\lr\right) \right]^t \phi''\left(\lr\right)
\]
and           (vi)           definition~\ref{ParametricPhiFisherDivergence:def},
eq.~\eqref{ParametricPhiFisherDivergence:def} of the nonparametric $\phi$-Fisher
matrix to obtain
\[
\opL_1\left(\nabla_\theta \Df \right)  + \opL_2\left(\nabla_\theta \Df \right) =
\int_\Omega    \!    \left[     \opK_1\left(\nabla_x    \left[    \beta_1    p_0
      \phi\left(\lr\right) \right] \right) +  \opK_2\left( \Hess_x \left[ p_0 \,
      \phi\left(\lr\right) \right] \right) \right] dx - \opK_2\left( \Jf \right)
+ \opL_2\left( \Jf_\theta \right)
\]

Again,  the vanishing  property  of $\beta_1  p_0  \phi\left(\lr\right)$ on  the
boundary of  $\Omega$ together with expression~\eqref{DivergenceTheoremNabla:eq}
of the divergence theorem, and  the vanishing assumption of $\nabla_x \left[ p_0
  \phi\left(\lr\right)\right]$     on     the     boundary,    together     with
expression~\eqref{DivergenceTheoremHessian:eq} of  the divergence theorem, allow
to see that the remaining integral term is zero, thus finishing the proof.


\section{Proof the generalized multivariate Guo's identities.}
\label{PhiGuo:app}

First of all,  note that if $f(M) =  g(x)$ with $x = M  u$, from $\frac{\partial
  f}{\partial m_{i,j}}  = \sum_k \frac{\partial  g}{\partial x_k} \frac{\partial
  x_k}{\partial  m_{i,j}}$ we  obtain $\nabla_M  f =  \left[ \nabla_x  g \right]
u^t$.

Now, $N$ being  independent of $X$ we  have $p_{Y|X=x}(y) = p_N( y  - \theta x)$
and from the expression of this Gaussian  law together with the fact that $R$ is
not parametrized by  $\theta$, one easilly shows that  $p_{Y|X=x}$ satisfies the
PDE
\begin{equation}
\left( \nabla_\theta \, p_{Y|X=x} \right) \theta^t = - \left( \nabla_y \left[y
\, p_{Y|X=x} \right] \right)^t - \left( \Hess_y \, p_{Y|X=x} \right) R \, .
\label{pde_pycond:eq}
\end{equation}
Note  now that  $\displaystyle p_Y(y)  = \int_\Omega  p_{Y|X=x}(y) \,  p_X(x) \,
dx$.  Following  the same  steps  that  in~\cite[Lemma~6.1]{Bar84}, detailed  in
appendix~\ref{HeatEquation:app}, one shows that  the gradient and the Hessian vs
$y$ and  the integration  in $x$ can  be interchanged  and that, provided  that $X$
admits a second  order moment\footnote{This requirement is not  necessary in the
  scalar  context.}, integration  vs $x$  and gradient  in $\theta$  can  also be
interchanged.    Thus,  multiplying   eq.~\eqref{pde_pycond:eq}  by   $p_X(x)$  and
integrating  over   $x$  allows   to  show  that   $p_Y$  satisfies   this  same
PDE~\eqref{pde_pycond:eq}.  This  PDE is of  the form~\eqref{pdegen_pdf:eq} with
$\opL_2 = 0$, $\opL_1(M) = M  \theta^t$, $\beta_1(y,\theta) = y$, $\opK_1(v) = -
v^t$    and    $\opK_2(M)    =    -    M   R$.     Hence,    immediately    from
proposition~\ref{PhiDivDeBruijn:prop}, we obtain
\begin{equation}
\left[ \nabla_\theta \Df(p_{Y|X=x} \| p_Y) \right] \theta^t = \Jf(p_{Y|X=x} \| p_Y) \, R
\label{PhiGuoConditional:eq}
\end{equation}
Now,  noting that  $\nabla_y \,  p_{Y|X=x}(y) =  - R^{-1}  \left( y  -  \theta x
\right) \, p_{Y|X=x}(y)$, one can also deduce that
\[
\nabla_y  \, p_Y(y)  =  - \int_\Omega  R^{-1} \left(  y  - \theta  x \right)  \,
p_{Y|X=x}(y) \, p_X(x) \, dx = - R^{-1} \left( y - \theta \Esp[X|Y=y] \right) \,
p_Y(y)
\]
leading to the following expression for the difference of the score functions
\[
\nabla_y \left[  \log\left( \frac{p_{Y|X=x}(y)}{p_Y(y)}\right) \right]  = R^{-1}
\theta \left( x - \Esp[X|Y=y] \right)
\]
and  thus
\[
\Jf(p_{Y|X=x} \|  p_Y) R =  R^{-1} \theta  \! \left( \!  \int_\Omega  \!\!  \left(  x -
    \Esp[X|Y\!\!=\!y]   \right) \! \left(   x   -  \Esp[X|Y\!\!=\!y]   \right)^t  \!    \left(
    \frac{p_{Y|X=x}(y)}{p_Y(y)}     \right)^{\!\!2}     \!\!      \phi''\!\left(
    \frac{p_{Y|X=x}(y)}{p_Y(y)} \right) \! p_Y(y) dy \right) \! \theta^t
\]
The  proof of  proposition~\ref{PhiGuo:prop} finishes  plugging  this expression
in~\eqref{PhiGuoConditional:eq},  again  noting  that  $\frac{p_{Y|X=x}}{p_Y}  =
\frac{p_{X,Y}}{p_X  p_Y}$  so that  $\displaystyle  \Df(p_{X,Y}  \|  p_X p_Y)  =
\int_\Omega  \Df(p_{Y|X=x} \|  p_Y) \,  p_X(x)  \, dx$,  multiplying both  sides
of~\eqref{PhiGuoConditional:eq} by $p_X(x)$ and integrating over $\Omega$.


\section{Various elements on the conditions needed in some of the propositions}
\label{Conditions:app}

The conditions set  in the propositions are used  to interchange derivation with
respect  to a  parameter and  integration, thanks  to the  dominated convergence
theorem and the mean value theorem.  Considering that the pdf and the considered
entropic functionals are sufficiently regular, these conditions are probably not
very  restrictive. In  what  follows, we  give  some elements  dealing with  the
Gaussian channel, that can serve as a guidance for more general situations.


\subsection{The  pdf of  the output  of the  Gaussian channel  follows  the heat
  equation.}
\label{HeatEquation:app}

It is shown  for instance in~\cite{Bar84, Bar86} in the scalar  case that pdf of
the output $Y  = X + \sqrt{\theta}  N$ of the Gaussian channel  follows the heat
equation.   The same  approach  naturally  applies when  $N$  is a  multivariate
Gaussian with covariance matrix $R$~\cite{Joh04}.  We recall here the main steps
in  the multivariate  context,  the scalar  one  being a  particular case.   The
principle consists  in writing  the output  pdf as a  convolution, to  derive it
versus the parameter or the  state, to interchange derivation and integrals, and
thus  to use  the heat  equation  followed by  the gaussian  pdf (channel  noise
pdf). The  very same steps are used  in the multivariate Cauchy  and can clearly
serves as a basis to treat the generalized case.

In what follows, to simplify the notations, we write
\[
\alpha = (2 \pi)^{-\frac{d}{2}} |R|^{-\frac12}  \qquad \mbox{and} \qquad u = (y-x)^t
R^{-1} (y-x) \ge 0
\]
so that the pdf of the  noise writes $p_N(y-x) = \alpha \, \theta^{-\frac{d}{2}}
\exp\left(-\frac{u}{2 \theta}\right)$.   We need then to be  able to interchange
derivation  vs $\theta$  and  the integration  that  gives the  output pdf,  and
similarly for integration and Hessian vs $y$.

\


\subsubsection{$\displaystyle   \frac{\partial}{\partial   \theta}   \int_\Omega
  p_X(x) p_N(y-x)  \, dx  = \int_\Omega p_X(x)  \frac{\partial}{\partial \theta}
  p_N(y-x) \, dx $}

A   direct  calculus   gives  $$\frac{\partial}{\partial   \theta}   p_N(y-x)  =
\frac{\alpha}{2} \, \theta^{-\frac{d}{2}-2} (u - \theta d) \exp\left(-\frac{u}{2
    \theta}\right).$$ A short study of this  function versus $u \ge 0$ allows to
prove  that  $\left|   \frac{\partial}{\partial  \theta}  p_N(y-x)  \right|  \le
\frac{\alpha \, d}{2 \, \theta^{\frac{d}{2}+1}}$ and thus that
\[
\left| \frac{\partial}{\partial \theta} p_N(y-x) p_X(x) \right| \le \frac{\alpha
  \, d}{2 \, \theta^{\frac{d}{2}+1}} p_X(x)
\]
which is  integrable. In other words  $\frac{\partial}{\partial \theta} p_N(y-x)
p_X(x)$ is $\theta$-locally dominated  in $\Rset_+^*$ by an integrable function,
which allows to conclude thanks to the dominated convergence theorem.

\


\subsubsection{$\displaystyle  \nabla_y  \int_\Omega  p_X(x)  p_N(y-x) \,  dx  =
  \int_\Omega p_X(x) \nabla_y p_N(y-x) \, dx $}

Direct algebra leads to
\[
\left\| R^{\frac12}  \nabla_y p_N(y-x) \right\|^2 = \alpha^2  \theta^{-d-2} u \,
\exp\left(- \frac{u}{\theta}\right)
\]
where $R^{\frac12}$  is the (unique) symmetric definite  positive matrix, square
root of  $R$. Studying this function  versus $u$, it is  straightforward to show
that
\[
\left\|     R^{\frac12}     \nabla_y     p_N(y-x)    \right\|     \le     \alpha
\operatorname{e}^{-\frac12} \theta^{-\frac{d+1}{2}}
\]
Thus, since $\nabla_y p_N(y-x) = R^{-\frac12} R^{\frac12} \nabla_x p_N(y-x)$ and
the  definition  of   the  matrix  2-norm~\cite{Mey00}\footnote{Note  that  from
  $\|\cdot\|_2 \le \|\cdot\|_F$~\cite{Mey00}, one  can replace the 2-norm by the
  Froebenius norm in the inequality.},
\[
\left\|  \nabla_y \,  p_N(y-x)  p_X(x) \right\|  \le \frac{\left\|  R^{-\frac12}
  \right\|_2}{\sqrt{\theta \operatorname{e}} \,  (2 \pi \theta)^{\frac{d}{2}} \,
  |R|^{\frac12}} \, p_X(x)
\]
which  is integrable.  Again,  $\nabla_y  p_X(x) p_N(y-x)$  is  dominated by  an
integrable function, allowing integration and derivation interchange.

\


\subsubsection{$\displaystyle  \Hess_y  \int_\Omega  p_X(x)  p_N(y-x) \,  dx  =
  \int_\Omega p_X(x) \Hess_y p_N(y-x) \, dx $}

Immediately from the pdf $p_N$ one has
\[
R^{\frac12}  \Hess_y  p_N(y-x)   R^{\frac12}  =  \alpha  \theta^{-\frac{d}{2}-2}
\exp\left(-  \frac{u}{2 \theta}\right) \left[  - \theta  I +  R^{-\frac12} (y-x)
  (y-x)^t R^{-\frac12} \right].
\]
Multiplying this expression by its  transposition and taking the trace to obtain
its Frobenius norm $\| \cdot \|_F$~\cite{Mey00}, one has thus
\[
\left\|  R^{\frac12}  \Hess_y   p_N(y-x)  R^{\frac12}  \right\|^2_F  =  \alpha^2
\theta^{-d-4}  \left(  u^2  -  2  \theta  u +  d  \theta^2  \right)  \exp\left(-
  \frac{u}{\theta}\right).
\]
A short study of this function vs $u \ge 0$ shows that,
\[
\left\| R^{\frac12} \, \Hess_y p_N(y-x)  \, R^{\frac12} \right\|_F \le \alpha \,
d^{\frac12} \theta^{-\frac{d}{2}-1}.
\]
From~\cite[p.~279]{Mey00} stating  that $\| A B \|_F  \le \| A \|_F  \| B \|_F$,
one obtains from  $\Hess_y p_N(y-x) = R^{-\frac12} \Big(  R^{\frac12} \, \Hess_y
p_N(y-x) \, R^{\frac12} \Big) R^{-\frac12}$
\[
\left\|  \Hess_y  p_X(x)  p_N(y-x)  \right\|_F  \le  \frac{\sqrt{d}  \,  \left\|
    R^{-\frac12}  \right\|_F^2}{\sqrt{\theta}  \,  (2 \pi  \theta)^{\frac{d}{2}}
  |R|^{\frac12}} \, p_X(x),
\]
which implies that $\Hess_y p_X(x)  p_N(y-x) = \nabla_y \left( \nabla_y^t p_X(x)
  p_N(y-x) \right)$  is dominated by  an integrable function, allowing  again to
finish the proof.


\subsection{Boundary conditions.}
\label{BoundaryConditions:app}


\subsubsection{$\phi(p_Y)$ vanishes in the boundary of $\Omega_Y$}

Note first that  $\Omega_Y = \Rset^d$ since $p_N > 0$  so that $p_X(x) p_N(y-x)
\ge 0$ cannot be identically zero. Furthermore,
\[
\left|  p_X(x)  p_N(y-x)  \right|  \le  p_X(x) \sup_{y  \in  \Rset^d}  p_N(y)  =
\frac{1}{(2 \pi \theta)^{\frac{d}{2}} |R|^{\frac12}} p_X(x)
\]
Hence, since $p_X(x)  p_N(y-x)$ is dominated by an  integrable function, one can
evoke the dominate convergence theorem to conclude that
\[
\lim_{\|y\| \to  \infty} p_Y(y) =  \int_{\Omega} p_X(x) \lim_{\|y\|  \to \infty}
p_N(y-x) \, dx = 0
\]
\ie $p_Y$ vanishes  in the boundary of $\Omega_Y$. Together  with $\phi(0) = 0$,
$\phi(p_Y)$ vanishes in the boundary of $\Omega_Y$.

\


\subsubsection{$\nabla_y \left[ \phi(p_Y) \right]$  vanishing in the boundary of
  $\Omega_Y$ under weak conditions}

Remind that we assume here that there exists a  \ $k \in (0 \; 1)$ \ such that \
$\lim_{u \to 0} u^k \phi'(u) =  0$.  This weak condition is sufficient to insure
the vanishing property of $\nabla_y \left[ \phi(p_Y) \right]$.

To show this, let us write
\[
\nabla_y \left[  \phi(p_Y) \right]  = \left[ \nabla_y  p_Y \right]  \phi'(p_Y) =
\frac{\nabla_y p_Y}{p_Y^k} p_Y^k \phi'(p_Y)
\]
From  the assumption  on  $\phi'$, since  $p_Y$ goes  to  0 in  the boundary  of
$\Omega_Y$,  the  quantity  $p_Y^k  \phi'(p_Y)$  vanishes  in  the  boundary  of
$\Omega_Y$.

Then,    from     $k    <    1$    one    applies     the    inverse    H\"older
inequality~\cite[th.~189]{HarLit52}  to   $\displaystyle  p_Y(y)  =  \int_\Omega
p_X(x) p_N(y-x) \, dx$ viewed as  a scalar product between $1$ and $p_N(y-x)$ of
kernel $p_X$, leading to
\[
p_Y(y)
\ge \left(  \int_\Omega p_X(x) \, dx  \right)^{\frac{1}{k^*}} \left( \int_\Omega
  p_X(x) \, \left[ p_N(y-x) \right]^k\, dx \right)^{\frac{1}{k}}
\]
where  $\frac{1}{k} +  \frac{1}{k^*}  = 1$  ($k^*$  is thus  negative). We  have
already seen that for $p_Y$ we can permute $\nabla_y$ and the integral, hence by
this permutation, bounding the norm of the integral by the integral of the norm,
and from the previous inequality, one obtains
\begin{eqnarray*}
\frac{\left\| \nabla_y p_Y(y) \right\|}{\left[ p_Y(y) \right]^k} & \le &
\frac{\displaystyle \int_\Omega p_X(x) \, \left\| \nabla_y p_N(y-x) \right\| \,
dx}{\displaystyle \int_\Omega p_X(x) \, \left[ p_N(y-x) \right]^k \,
dx}\\[2.5mm]
& = & \frac{\displaystyle \int_\Omega p_X(x) \, \left[ p_N(y-x) \right]^k
\left\| \frac{\nabla_y p_N(y-x)}{\left[ p_N(y-x) \right]^k} \right\| \,
dx}{\displaystyle \int_\Omega p_X(x) \, \left[ p_N(y-x) \right]^k \,
dx}\\[2.5mm]
& \le & \sup_{x \in \Omega} \left\| \frac{\nabla_y p_N(y-x)}{\left[ p_N(y-x)
\right]^k} \right\|\\[2.5mm]
& \le & \sup_{y \in \Rset^d} \left\| \frac{\nabla_y p_N(y)}{\left[ p_N(y)
\right]^k} \right\|
\end{eqnarray*}

Now, a rapid study of
\[
\left\| \frac{R^{\frac12} \nabla_y  p_N(y) }{\left[ p_N(y) \right]^k} \right\|^2
=  \alpha^{2(1-k)}  \theta^{-(1-k) d  -  2} \,  u  \,  \exp\left( -  \frac{(1-k)
    u}{\theta}\right)
\]
versus $u$ allows to show that
\[
\left\| \frac{\nabla_y p_N(y)}{\left[ p_N(y) \right]^k} \right\| \le
\frac{\left\| R^{-\frac12} \right\|_2}{\sqrt{(1-k) \operatorname{e} \theta} \,
(2 \pi \theta)^{\frac{(1-k) d}{2}} |R|^{\frac{1-k}{2}}}
\]
As   a  conclusion,   $\sup_{y  \in   \Rset^d}  \frac{\left\|   \nabla_y  p_N(y)
   \right\|}{\left[ p_N(y) \right]^k}$ is finite, which finishes the proof.


\bibliography{DeBruijnGeneralization}
\bibliographystyle{unsrt}

\end{document}